\documentclass[twocolumn]{aastex631}
\usepackage{enumerate}
\usepackage{amsmath,amsthm,amsfonts,amssymb,amscd}
\usepackage{threeparttable,booktabs}

\received{XXX XX, XXXX}
\revised{XXX XX, XXXX}
\accepted{XXX XX, XXXX}
\submitjournal{ApJ}

\shorttitle{PAC. IV. Percent precision constraints on SHMR}
\shortauthors{Xu et al.}

\begin{document}
\defcitealias{2022ApJ...925...31X}{Paper I}
\defcitealias{2022ApJ...939..104X}{Paper III}
\defcitealias{2013ApJ...770...57B}{BP13}
\title{Photometric Objects Around Cosmic Webs (PAC) Delineated in a Spectroscopic Survey. IV.  High Precision Constraints on the Evolution of Stellar-Halo Mass Relation at Redshift $z<0.7$}

\correspondingauthor{Y.P. Jing}
\email{ypjing@sjtu.edu.cn}

\author[0000-0002-7697-3306]{Kun Xu}
\affil{Department of Astronomy, School of Physics and Astronomy, Shanghai Jiao Tong University, Shanghai, 200240, People’s Republic of China}

\author[0000-0002-4534-3125]{Y.P. Jing}
\affil{Department of Astronomy, School of Physics and Astronomy, Shanghai Jiao Tong University, Shanghai, 200240, People’s Republic of China}
\affil{Tsung-Dao Lee Institute, and Shanghai Key Laboratory for Particle Physics and Cosmology, Shanghai Jiao Tong University, Shanghai, 200240, People’s Republic of China}

\author[0000-0001-6575-0142]{Yun Zheng}
\affiliation{Department of Astronomy, School of Physics and Astronomy, Shanghai Jiao Tong University, Shanghai, 200240, People’s Republic of China}

\author{Hongyu Gao}
\affil{Department of Astronomy, School of Physics and Astronomy, Shanghai Jiao Tong University, Shanghai, 200240, People’s Republic of China}



\begin{abstract}
Taking advantage of the Photometric objects Around Cosmic webs (PAC) method developed in Paper I, we measure the excess surface density $\bar{n}_2w_{{\rm{p}}}$ of photometric objects around spectroscopic objects down to stellar mass $10^{8.0}M_{\odot}$, $10^{9.2}M_{\odot}$ and $10^{9.8}M_{\odot}$ in the redshift ranges of $z_s$\footnote{Throughout the paper, we use $z_s$ for spectroscopic redshift, $z$ for the $z$-band magnitude.}$<0.2$, $0.2<z_s<0.4$ and $0.5<z_s<0.7$ respectively, using the data from the DESI Legacy Imaging Surveys and the spectroscopic samples of Slogan Digital Sky Survey (i.e. Main, LOWZ and CMASS samples). We model the measured $\bar{n}_2w_{{\rm{p}}}$ in N-body simulation using abundance matching method and constrain the stellar-halo mass relations (SHMR) in the three redshift ranges to percent level. With the accurate modeling, we demonstrate that the stellar mass scatter for given halo mass is nearly a constant, and that the empirical form of Behroozi et al describes the SHMR better than the double power law form at low mass. Our SHMR accurately captures the downsizing of massive galaxies since $z_s=0.7$, while it also indicates that small galaxies are still growing faster than their host halos. The galaxy stellar mass functions (GSMF) from our modeling are in perfect agreement with the {\it model-independent} measurements in Paper III, though the current work extends the GSMF to a much smaller stellar mass. Based on the GSMF and SHMR, we derive the stellar mass completeness and halo occupation distributions for the LOWZ and CMASS samples, which are useful for correctly interpreting their cosmological measurements such as galaxy-galaxy lensing and redshift space distortion. 
\end{abstract}

\section{Introduction} \label{sec:intro}
With the rapid development of cosmology, we have entered an era of understanding galaxy formation in the cosmological framework \citep{2010gfe..book.....M,2012AnP...524..507F,2015ARA&A..53...51S, 2017ARA&A..55...59N}. In this framework, galaxies reside in the dark matter halos, and the growth of galaxies is closely related to the growth of their host halos. Thus, precise measurement of the galaxy-halo connection is one of the most important issues in galaxy formation \citep{2018ARA&A..56..435W}. Moreover, accurate constraints on the the galaxy-halo connection can in turn benefit the cosmology studies, such as galaxy-galaxy gravitational lensing \citep{2001PhR...340..291B,2010ARA&A..48...87T} and redshift space distortion of galaxy distribution \citep{1987MNRAS.227....1K,2004PhRvD..70h3007S}, by connecting the observations of galaxies to the theories of dark matter halos.

The stellar-halo mass relation (SHMR) is one of the most commonly used relations to populate galaxies to halos, in which larger halos host more massive galaxies with a relatively tight scatter. Various methods have been used to model the SHMR at different redshifts, including abundance matching \citep[AM,][]{2010MNRAS.404.1111G,2010MNRAS.402.1796W}, conditional luminosity function \citep[CLF,][]{2012ApJ...752...41Y} and empirical modeling \citep[EM,][]{2013MNRAS.428.3121M,2018MNRAS.477.1822M,2013ApJ...770...57B,2019MNRAS.488.3143B}. No matter how the methods were implemented, they all rely on the measurements of observables such as the galaxy stellar mass function (GSMF) and galaxy clustering (GC). Although relatively tight constraints have achieved at the high mass end of the SHMR for local $z_s\approx 0$, the results at the low mass end still vary widely between studies \citep[e.g.][]{2010MNRAS.404.1111G,2012ApJ...752...41Y,2013MNRAS.428.3121M,2019MNRAS.488.3143B}
, due to the lack of accurate measurements of GSMF and GC at the faint end. At higher redshift, the measurement of SHMR is more difficult and more uncertain since there does not exist a large stellar-mass limited redshift sample. 

Measurements of GSMF and GC rely on accurate redshift
information. In the past two decades, there has been significant
progress in spectroscopic surveys \citep{2000AJ....120.1579Y,2001MNRAS.328.1039C, 2003ApJ...592..728S, 2005A&A...439..845L, 2012ApJS..203...21A, 2012AJ....144..144B, 2014A&A...562A..23G, 2014PASJ...66R...1T, 2016arXiv161100036D, 2020ApJS..249....3A}. In the local universe ($z_s\sim0$), the measurements of GSMF and GC are mainly from the large spectroscopic surveys, in particular the Sloan Digital Sky Survey \citep[SDSS,][]{2000AJ....120.1579Y} and the Two Degree
Field Galaxy spectroscopic survey \citep[2dFGRS,][]{2001MNRAS.328.1039C}), down to $10^{9.0}M_{\odot}$  \citep{2001MNRAS.326..255C,2002MNRAS.332..827N,2006MNRAS.368...21L,2008MNRAS.388..945B,2009MNRAS.398.2177L}. However, for the faint galaxies, the accuracy of the measurements, especially the GC, is still very limited by the survey volume. At intermediate redshifts ($z_s<1.0$), deeper spectroscopic surveys begin to dominate the measurements. The DEEP2 Galaxy spectroscopic survey \citep{2003SPIE.4834..161D}, the VIMOS-VLT Deep Survey \citep[VVDS,][]{2005A&A...439..845L} and the VIMOS Public Extragalactic spectroscopic survey \citep[VIPERS,][]{2014A&A...562A..23G} have been used to successfully measure GC and GSMF for galaxies down to $10^{10.0} M_{\odot}$ \citep{2007A&A...474..443P,2008A&A...478..299M,2013ApJ...767...89M,2013A&A...557A..17M,2013A&A...558A..23D}, though the measurements are limited by the survey volumes at the high stellar mass end (i.e. $>10^{11.0} M_{\odot}$). Similarly, measuring GC and GSMF at the faint end (i.e. $<10^{10.0} M_{\odot}$) is also very challenging at $z_s\sim 0.5$ for the magnitude-limited redshift surveys.

Many attempts have been made to push the measurement of GSMF to higher redshifts ($z_s>1.0$) using deep multi-band photometric surveys \citep{2006A&A...459..745F,2013A&A...556A..55I,2013ApJ...777...18M,2014ApJ...783...85T,2015MNRAS.447....2M,2017A&A...605A..70D,2018MNRAS.480.3491W,2020ApJ...893..111L,2021MNRAS.503.4413M,2022A&A...664A..61S}, which are usually deeper and more complete in terms of the stellar mass. However, the deep multi-band photometric surveys usually have very small survey area ($1\sim2\ \rm{deg}^2$), since they require long exposure time to reach very faint sources and multiple bands from UV to IR to obtain relatively accurate photometric redshifts (photoz) for faint objects. Thus, the cosmic variance could be significant, especially for massive galaxies. In addition, although faint sources are detected in the deep photometric surveys, the photometric redshifts derived for them are usually trained with the the spectroscopic data, which may not be complete for all type of faint sources, and therefore lead to larger photoz errors. These effects may all introduce uncertainties in the measurements of GSMF and GC.

Systematic bias between the measurements in different surveys can also affect the study of SHMR. Many studies \citep{2012ApJ...752...41Y,2013MNRAS.428.3121M,2018MNRAS.477.1822M,2013ApJ...770...57B,2019MNRAS.488.3143B} collected the GSMF measurements from different surveys at different redshifts to model the evolution of SHMR. However, as shown in \citet{2022ApJ...939..104X}, the systematic bias of GSMF is still very large ($\sim30\%$) between different studies at the high mass end even after careful calibration, which may significantly influence the results of SHMR.

In \citet[hearafter \citetalias{2022ApJ...925...31X}]{2022ApJ...925...31X}, on the basis of \citet{2011ApJ...734...88W}, we developed a method named Photometric objects Around Cosmic webs (PAC) to estimate the excess surface density $\bar{n}_2w_{{\rm{p}}}$ of photometric objects with certain physical properties around spectroscopically identified sources, which can take full use of the spectroscopic and deeper photometric surveys. With PAC, we can measure $\bar{n}_2w_{{\rm{p}}}$ for galaxies to a much lower stellar mass than one can with the spectroscopic survey only.  Another advantage is that  $\bar{n}_2w_{{\rm{p}}}$ is measured in a uniform way at different redshifts, since the same photometric catalog and the same analysis method (i.e stellar mass measurement method) are used. Obviously the quantity contains the desired information of both GSMF and GC.  

In this paper, we will measure $\bar{n}_2w_{{\rm{p}}}$ at different redshifts with Dark Energy Camera Legacy Survey (DECaLS) photometric catalogs and Slogan Digital Sky Survey (SDSS) Main, LOWZ and CMASS spectroscopic samples. Then, we will model the PAC measurements using the abundance matching method and constrain the evolution of the SHMR. We will compare the derived GSMF with the model independent measurements from \citet[hearafter \citetalias{2022ApJ...939..104X}]{2022ApJ...939..104X} to test the reliability of the results, and we will show how much we can extend GSMF to a small stellar mass in this study. Based on our SHMR and GSMF, we will also derive the stellar mass completeness and halo occupation distributions \citep[HOD,][]{1998ApJ...494....1J,2000MNRAS.318.1144P,2000ApJ...543..503M,2000MNRAS.318..203S,2002ApJ...575..587B,2003MNRAS.339.1057Y,2005ApJ...633..791Z,2015MNRAS.454.1161Z} for the LOWZ and CMASS samples, which can be useful for cosmological interpretation of their clustering and lensing measurements.

We introduce the PAC method and show the measurements in Section \ref{sec:obervation}. In Section \ref{sec:simulation}, we model the measurements in N-body simulation and give the SHMR. In Section \ref{sec:hod}, we derive the HODs of spectroscopic samples. We briefly summarize our results in Section \ref{sec:sum}. We adopt the cosmology with $\Omega_m = 0.268$, $\Omega_{\Lambda} = 0.732$ and $H_0 = 71{\rm \ km/s/Mpc}$ throughout the paper.

\section{observation and measurements} \label{sec:obervation}
In this section, we first give a brief summary about the Photometric objects Around Cosmic webs (PAC) method and the spectroscopic and photometric samples used in this work. Then we investigate the stellar mass completeness of the samples and present the final PAC measurements.

\subsection{Photometric objects Around Cosmic webs (PAC)}
Supposing we want to study two populations of galaxies, with one (${\rm{pop}}_1$) from a spectroscopic catalog and the other (${\rm{pop}}_2$) from a photometric catalog, within a relatively narrow redshift range. We proposed a method called PAC in \citetalias{2022ApJ...925...31X} that can accurately measure the excess surface density $\bar{n}_2w_{\rm{p}}(r_{\rm{p}})$ of ${\rm{pop}}_2$ around ${\rm{pop}}_1$ with certain physical properties:
\begin{equation}
    \bar{n}_2w_{\rm{p}}(r_{\rm{p}}) = \frac{\bar{S}_2}{r_1^2}w_{12,\rm{weight}}(\theta)\,\,,\label{eq:1}
\end{equation}
where $\bar{n}_2$ and $\bar{S}_2$ are the mean number density and mean angular surface density of $\rm{pop}_2$, $r_1$ is the comoving distance to $\rm{pop}_1$, and $w_{\rm{p}}(r_{\rm{p}})$ and $w_{12,\rm{weight}}(\theta)$ are the projected cross-correlation function (PCCF) and the weighted angular cross-correlation function (ACCF) between $\rm{pop}_1$ and $\rm{pop}_2$ with $r_{\rm{p}}=r_1\theta$. Since $\rm{pop}_1$ has a redshift distribution, $w_{12}(\theta)$ is weighted by $1/r^2_1$ to account for the effect that, at fixed $\theta$, $r_{\rm{p}}$ varies with $r_1$. Using PAC, we can measure the rest-frame physical properties of $\rm{pop}_2$ statistically without the need of the redshift information of $\rm{pop}_2$, for which we can take full use of the deep photometric surveys. The following are the main steps of PAC:
\begin{enumerate}[(i)]
    \item Split $\rm{pop}_1$ into narrower redshift bins, mainly accounting for the fast change of $r_1$ with redshift.
    \item Assuming all galaxies in $\rm{pop}_2$ have the same redshift as the mean redshift of each redshift bin, calculate the physical properties of $\rm{pop}_2$ using methods such as spectral energy distribution (SED). Therefore, in each redshift bin of $\rm{pop}_1$, there is a physical property catalog of $\rm{pop}_2$ .
    \item In each redshift bin, select $\rm{pop}_2$ with certain physical properties and calculate $\bar{n}_2w_{\rm{p}}(r_{\rm{p}})$ according to Equation \ref{eq:1}. The foreground and background objects with wrong properties are cancelled out through ACCF and only $\rm{pop}_2$ around $\rm{pop}_1$ with correct redshifts left.
    \item Combine the results from different redshift bins by averaging with proper weights.
\end{enumerate}
For more details, we refer to \citetalias{2022ApJ...925...31X}.

\begin{table*}
    \centering
    \caption{Final designs for the PAC measurements.}
    \begin{threeparttable}
    \setlength{\tabcolsep}{6mm}{
    \begin{tabular*}{\hsize}{ccccccc}
     \toprule
     redshift & Survey & $\rm{pop}_1$\tnote{a} & $\rm{pop}_2$\tnote{b} & PAC redshift bins \\
      &  & ($M_{\odot}$) & ($M_{\odot}$) &  \\
     \midrule
     $[0.05,0.2]$ & Main & $[10^{10.3},10^{11.3}]$ & $[10^{7.9},10^{11.7}]$ & $[0.05,0.075],[0.075,0.1],[0.1,0.15],[0.15,0.2]$\\
     $[0.2,0.4]$ & LOWZ & $[10^{11.3},10^{11.9}]$ & $[10^{9.1},10^{11.9}]$ & $[0.2,0.3],[0.3,0.4]$ \\
     $[0.5,0.7]$ & CMASS & $[10^{11.3},10^{11.9}]$ & $[10^{9.7},10^{11.9}]$ & $[0.5,0.6],[0.6,0.7]$ \\
     \bottomrule
    \end{tabular*} }
    \begin{tablenotes}
     \footnotesize
     \item[a] Stellar mass ranges of $\rm{pop}_1$ with an fiducial equal logarithmic bin width of $10^{0.2}M_{\odot}$.
     \item[b] Stellar mass ranges of $\rm{pop}_2$ with an fiducial equal logarithmic bin width of $10^{0.2}M_{\odot}$.
    \end{tablenotes}
    \end{threeparttable}
    \label{tab:t1}
\end{table*}

\begin{figure*}
    \plottwo{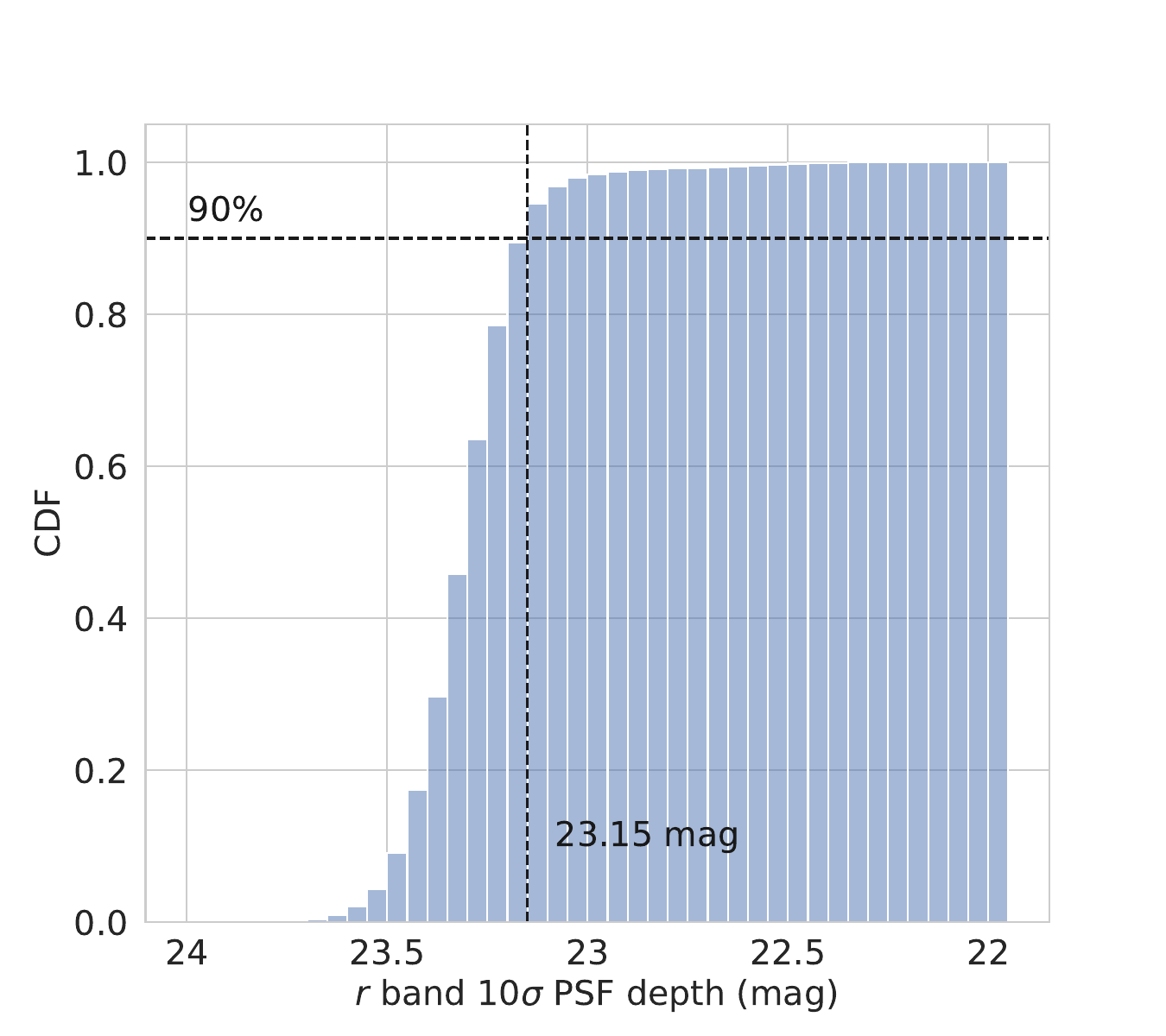}{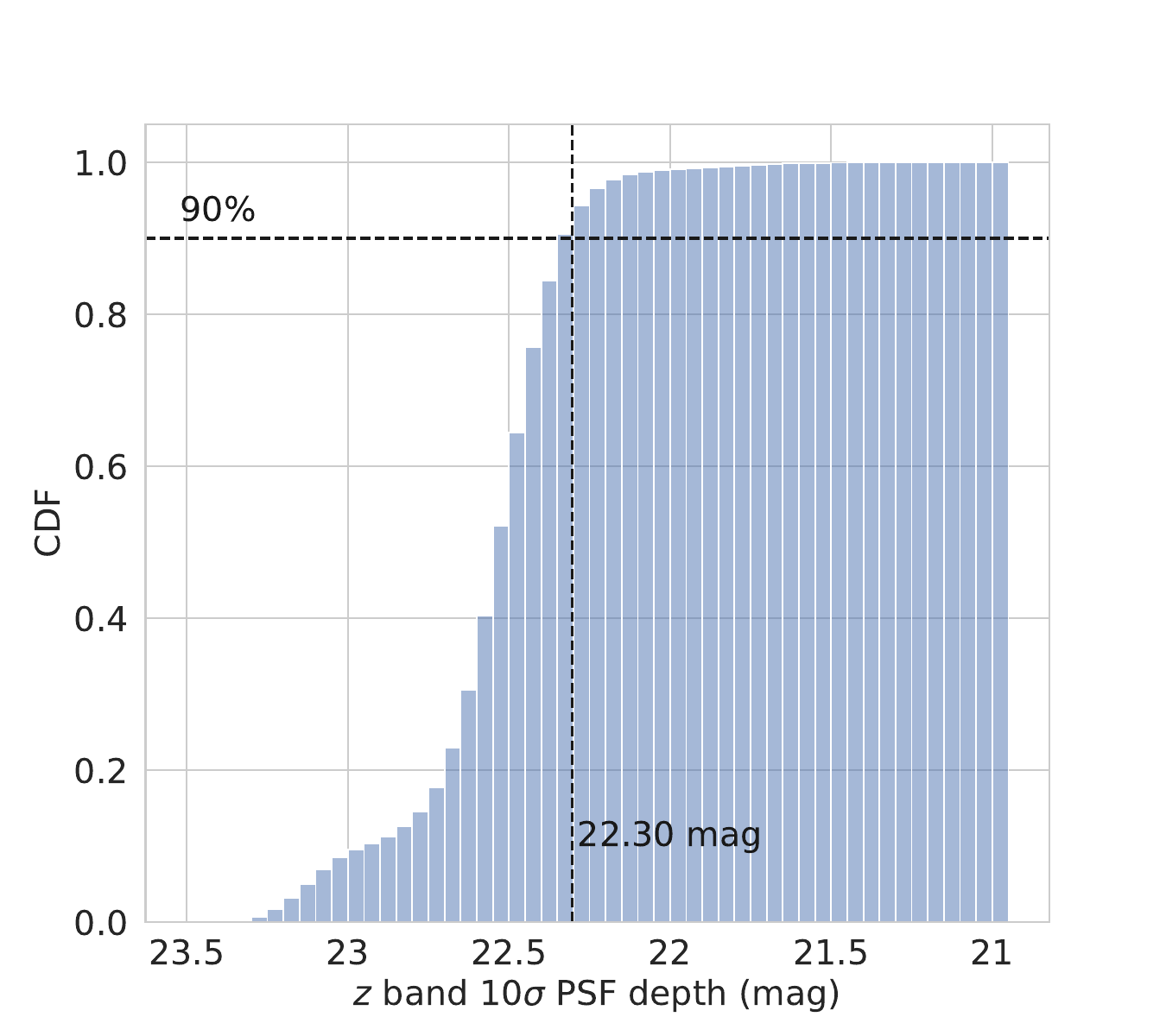}
    \caption{Cumulative distribution functions of the $r$ (left) and $z$ (right) band $10\sigma$ point source depths for DECaLS.}
    \label{fig:fig1}
\end{figure*}

\begin{figure*}[t]
    \plotone{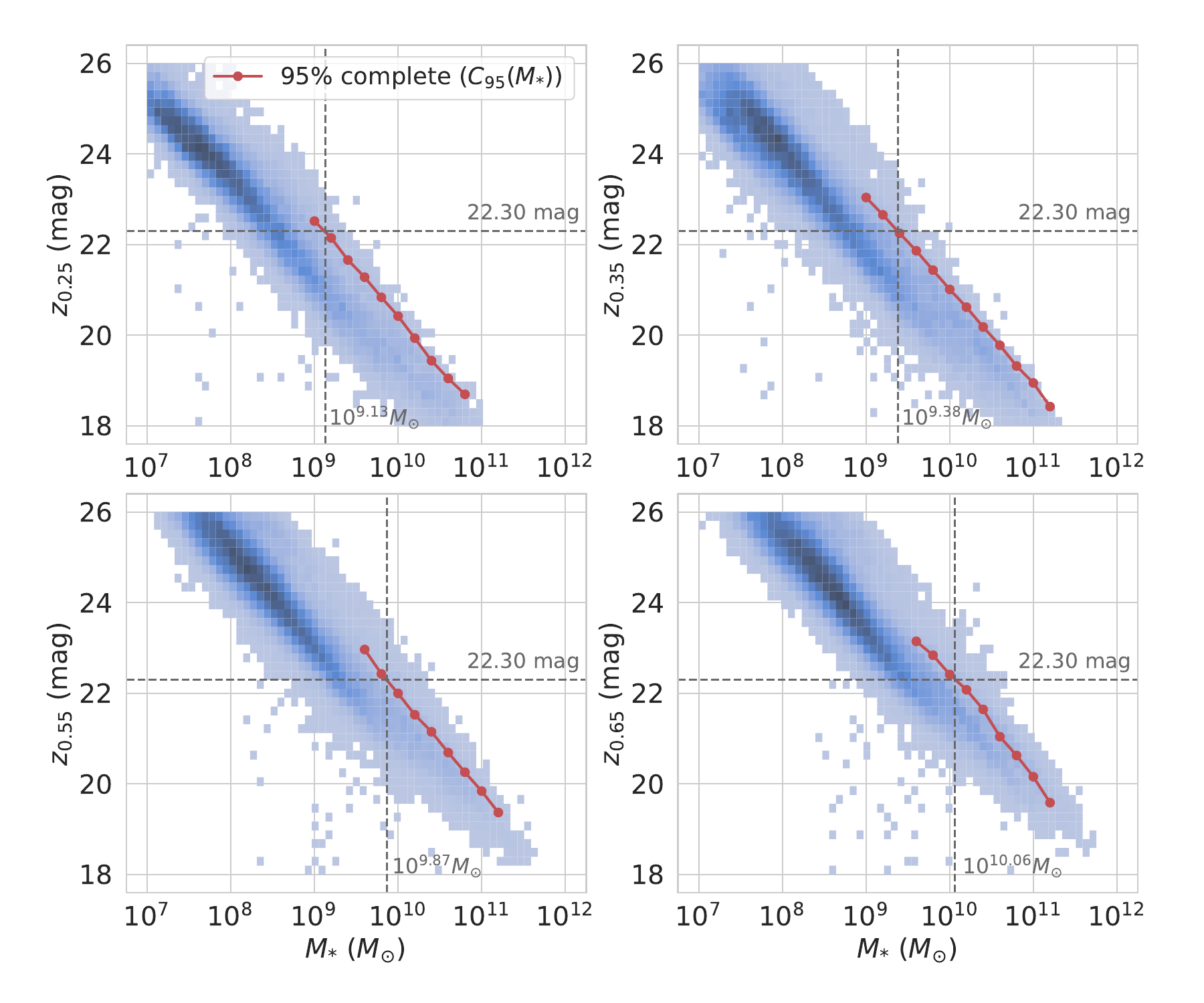}
    \caption{Stellar mass - $z$ band magnitude relations for DES Deep Field galaxies in redshift ranges of $[0.2,0.3]$, $[0.3,0.4]$, $[0.5,0.6]$ and $[0.6,0.7]$. Red lines with dots show the $95\%$  completeness limits $C_{95}(M_{*})$ varying with the stellar mass at each redshifts. Grey dashed lines show the stellar mass limits for DECLaS with the $z$ band depth of $22.30\ \rm{mag}$.}
    \label{fig:fig2}
\end{figure*}

\begin{figure*}[t]
    \plotone{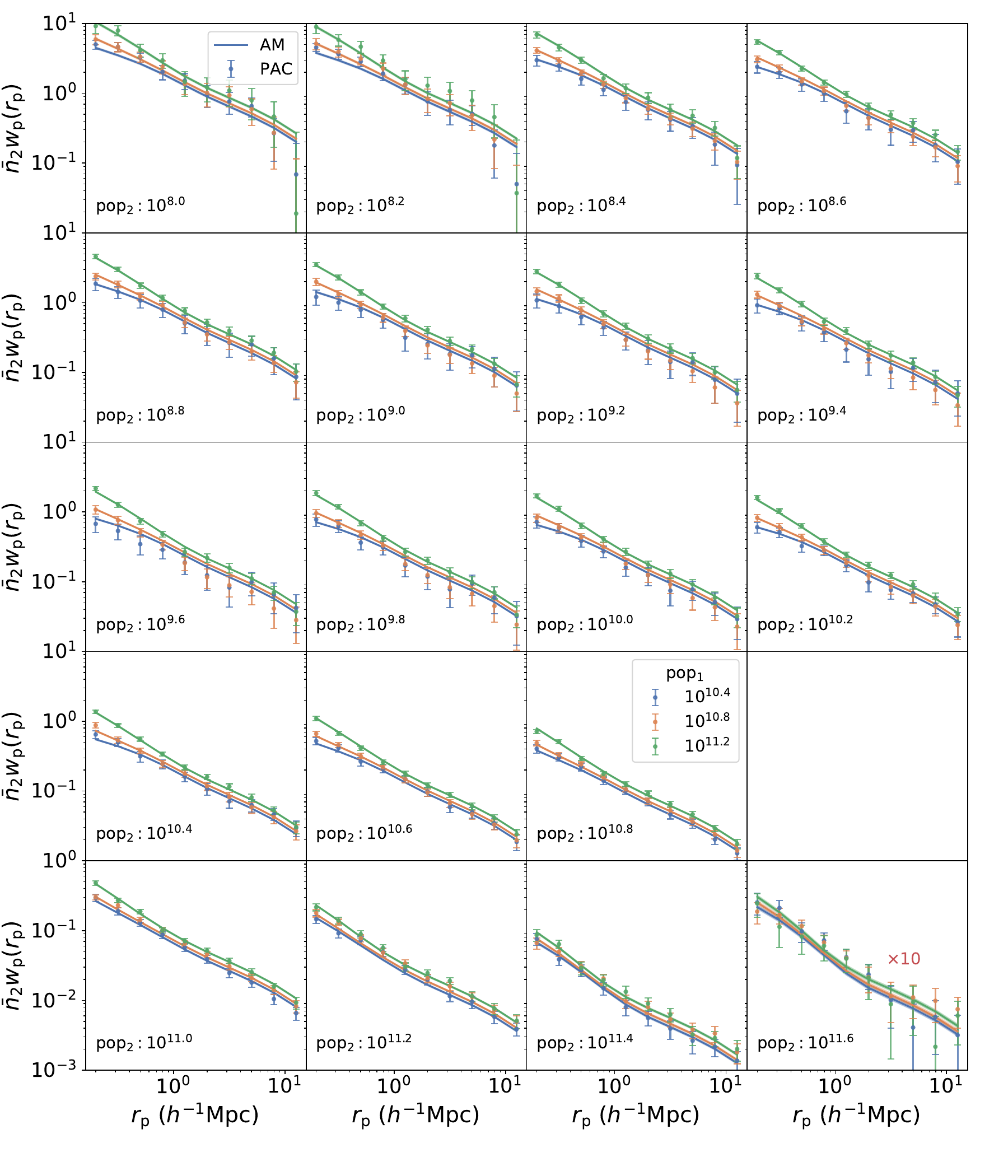}
    \caption{PAC measurements and fittings from the \citetalias{2013ApJ...770...57B} model for the Main sample redshift range ($z_s<0.2$) according to the designs in Table \ref{tab:t1}. Dots with error bars show the measurements and lines with shadows are the best-fit results and $1\sigma$ errors from AM modelings. We only show the results of 3 ${\rm{pop}}_1$ mass bins ($10^{10.4}, 10^{10.8}, 10^{11.2}M_{\odot}$) out of the total 5 for better illustration, and the results of the last ${\rm{pop}}_2$ bins ($10^{11.6}M_{\odot}$) are multiplied by $10$.}
    \label{fig:fig3}
\end{figure*}
 
 \begin{figure*}[t]
    \plotone{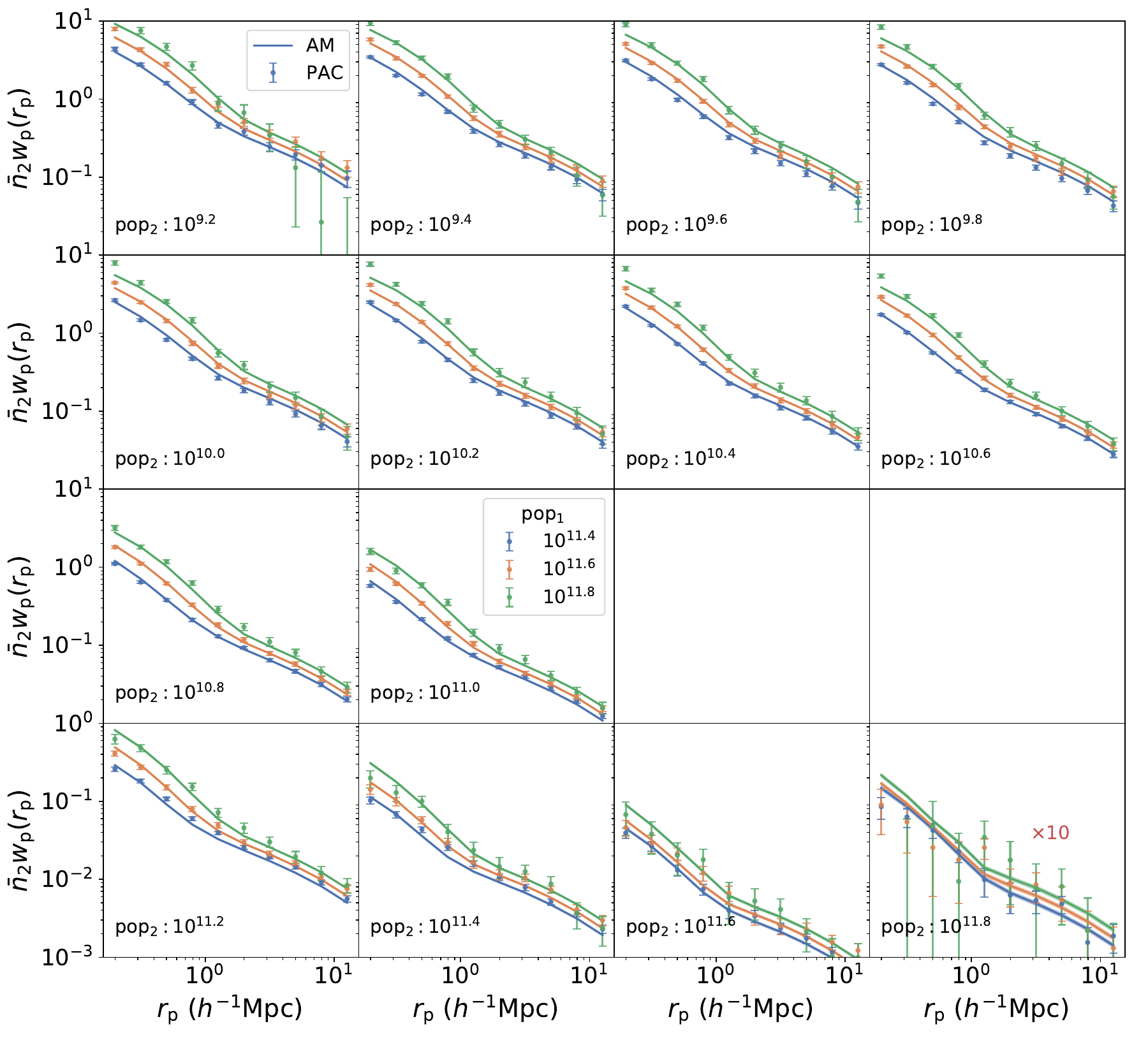}
    \caption{The same as Figure \ref{fig:fig3} but for the LOWZ redshift range ($0.2<z_s<0.4$).}
    \label{fig:fig4}
\end{figure*}
 
 \begin{figure*}[t]
    \plotone{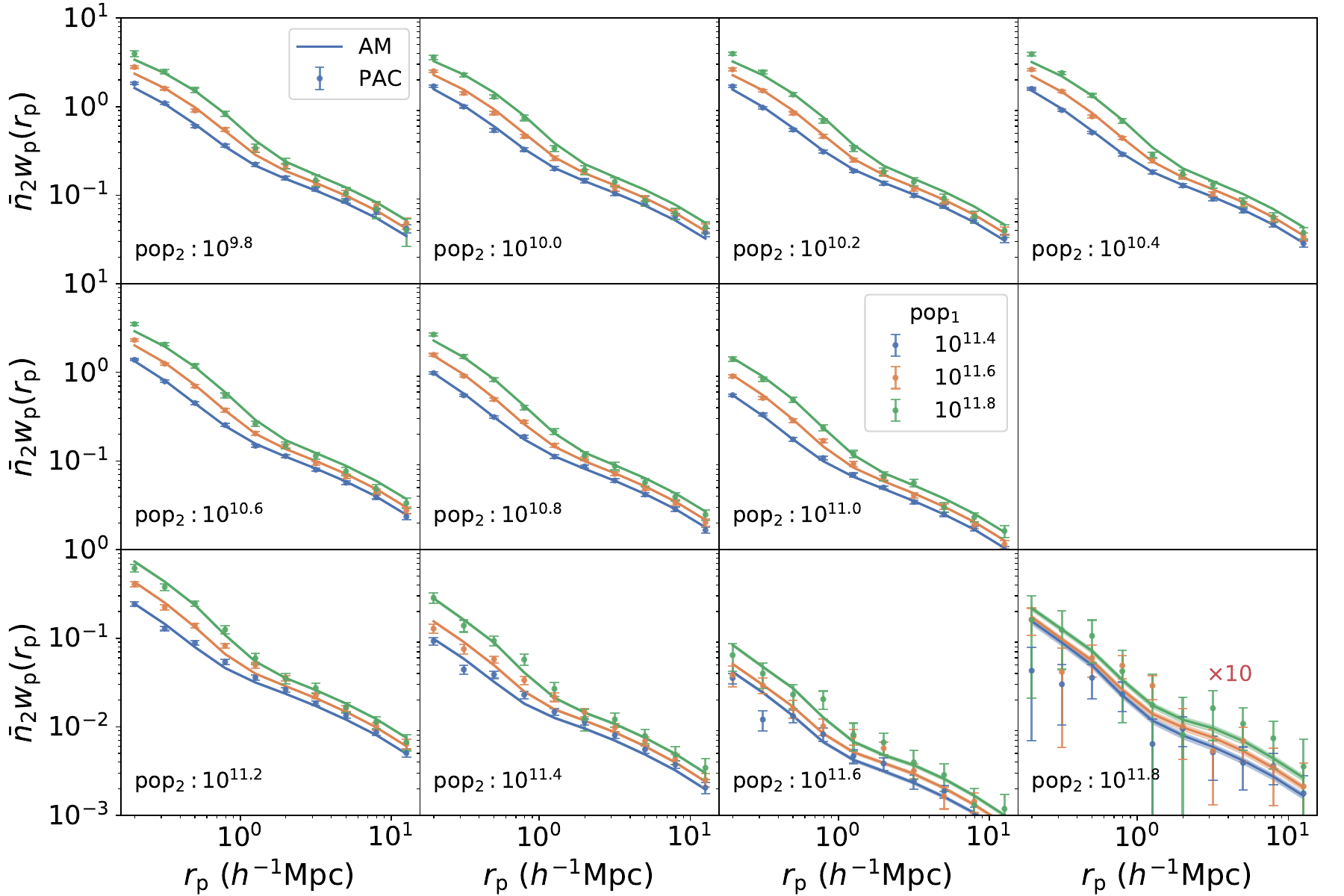}
    \caption{The same as Figure \ref{fig:fig3} but for the CMASS redshift range ($0.5<z_s<0.7$).}
    \label{fig:fig5}
\end{figure*}

\begin{figure*}[t]
    \plotone{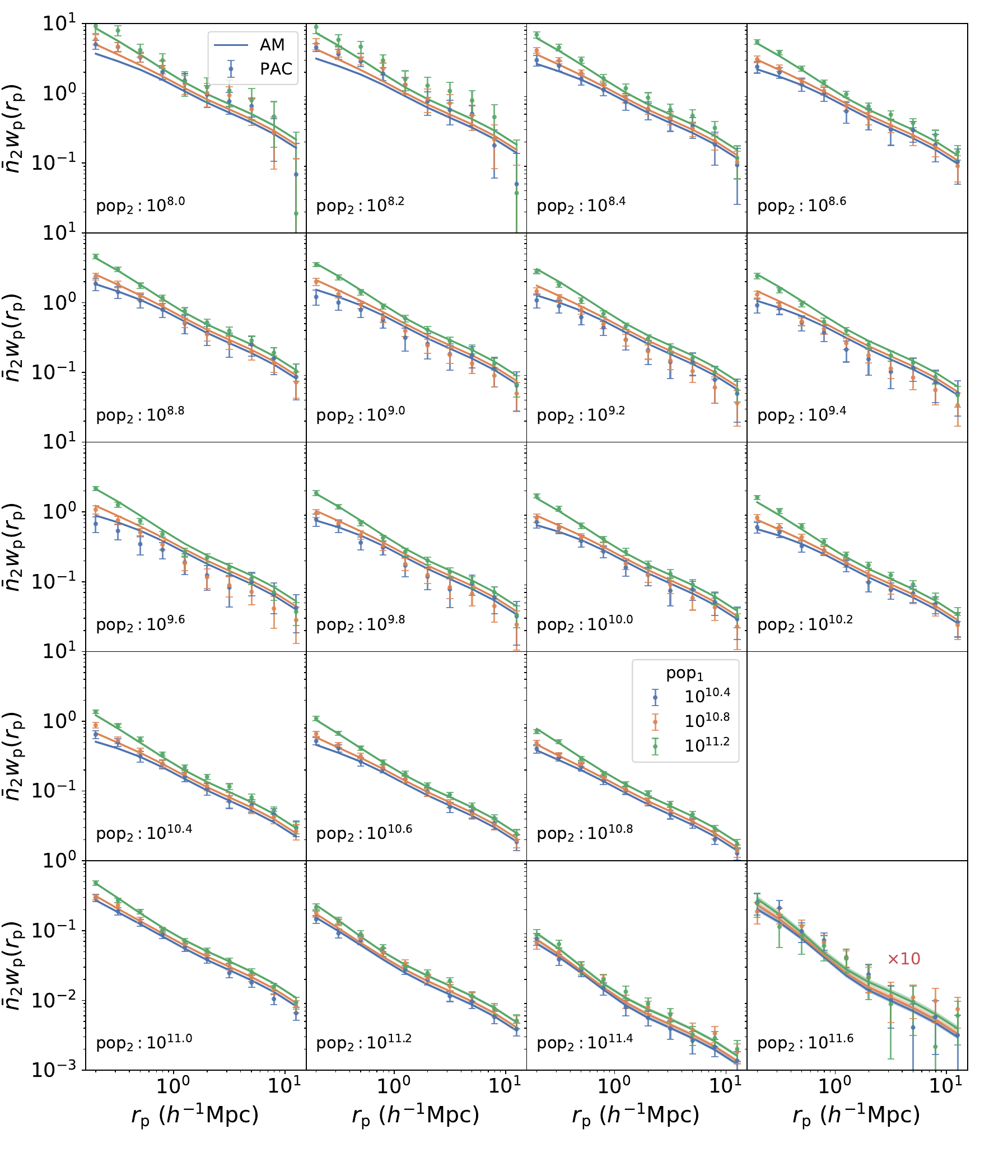}
    \caption{The same as Figure \ref{fig:fig3} but for the Main sample redshift range ($z_s<0.2$) with fittings from the DP model.}
    \label{fig:figz1}
\end{figure*}

\subsection{Spectroscopic and photometric samples}
In this work, we use the same spectroscopic and photometric data as used in \citetalias{2022ApJ...939..104X}. We summary their key features in the following.

For photometric data, we use photometric catalog\footnote{https://www.legacysurvey.org/dr9/catalogs/} of DECaLS from the DR9 of the DESI Legacy Imaging Survey \citep{2019AJ....157..168D}. It observes around $9000\ {\rm{deg}}^2$ in both the Northern and Southern Galactic caps (NGC and SGC) at $\rm{Dec}\leq32\ \rm{deg}$ in $g$, $r$ and $z$ bands with median $5\sigma$ point source depths of $24.9$, $24.2$ and $23.3$ respectively. DECaLS also includes the data from the deeper Dark Energy Survey (DES; \citealt{2016MNRAS.460.1270D}) covering additional $5000\ {\rm{deg}}^2$ in the SGC. The images are processed using {\texttt{Tractor}}\footnote{https://github.com/dstndstn/tractor} \citep{2016ascl.soft04008L} to perform source extraction. The sources are then modeled with parametric profiles convolved with a specific point spread function (PSF), including a delta function for the point source, exponential law, de Vaucouleurs law,  and a S\'ersic profile. We use their best-fit model magnitudes throughout the paper. We only use the footprints that have been observed at least once in all three bands, and perform bright star mask and bad pixel mask to the catalog using the MASKBITS\footnote{https://www.legacysurvey.org/dr9/bitmasks/} provided by the Legacy Surveys. Additional masks are used to match the geometry of the spectroscopic sample in each redshift. Galactic extinction is corrected for all the sources using the maps of \citet{1998ApJ...500..525S}. To reject stars, we exclude sources with point source (PSF) morphologies, and we adopt color cuts in $r-z$ vs. $z-W1$ diagram calibrated in \citetalias{2022ApJ...939..104X} (see Figure 1) to further remove stellar objects, with $W1$ band data from the Wide-field Infrared Survey Explorer (WISE) \citep{2010AJ....140.1868W}. The stars have
\begin{equation}
    (z-W1<0.8\times(r-z)-1.0)\ {\rm{AND}}\ (r-z>1.0)\,\,.
\end{equation}

For spectroscopic data, all the catalogs used in this work are from SDSS \citep{2000AJ....120.1579Y}). We use the SDSS DR7 Main sample\footnote{http://sdss.physics.nyu.edu/lss/dr72/bright/} \citep{2009ApJS..182..543A}, SDSS-III BOSS DR12 LOWZ and CMASS samples\footnote{https://data.sdss.org/sas/dr12/boss/lss/} \citep{2015ApJS..219...12A,2016MNRAS.455.1553R} for three redshift ranges $z_s<0.2$, $0.2<z_s<0.4$ and $0.5<z_s<0.7$ respectively. All the three samples are selected with $\rm{Dec}\leq32\ \rm{deg}$ to match the footprint of DECaLS, and for the Main sample, only the NGC part is used since the SGC part is very small ($532\ {\rm{deg}}^2$). The spectroscopic sources are cross-matched with DECaLS to get the $g$, $r$ and $z$ band flux measurements. 

\subsection{Completeness and Designs}
Based on the test in \citetalias[see Figure 3 there]{2022ApJ...925...31X}, we split LOWZ and CMASS samples into 2 and Main sample into 4 narrower redshift bins as listed in Table \ref{tab:t1}. According to the stellar mass distributions (\citetalias[Figure 2]{2022ApJ...939..104X}), we choose the stellar mass range of $\rm{pop}_1$ as $[10^{10.3},10^{11.3}]M_{\odot}$ for the Main sample and $[10^{11.3},10^{11.9}]M_{\odot}$ for LOWZ and CMASS, and we further split them into several stellar mass bins with an equal logarithmic interval of $0.2$. 

As in \citetalias{2022ApJ...925...31X} and \citetalias{2022ApJ...939..104X}, we use the $r$ (for $z_s<0.2$) and $z$ (for $0.2<z_s<0.4$ and $0.5<z_s<0.7$) band $10\sigma$ point source depths of DECaLS to study the mass completeness of $\rm{pop}_2$. As shown in Figure \ref{fig:fig1}, $90\%$ of the regions in DECaLS are deeper than $23.15$ mag and $22.30$ mag in $r$ band and $z$ band respectively. Thus, we use $r=23.15$ and $z=22.30$ as the galaxy depths for DECaLS.

In \citetalias{2022ApJ...939..104X} (see Figure 3), using the Galaxy And Mass Assembly (GAMA) DR4 spectroscopic data \citep{2022MNRAS.513..439D}, we find that for $r$ band galaxy depth of $23.15\ \rm{mag}$, the complete stellar masses are $10^{7.61}M_{\odot}$, $10^{7.89}M_{\odot}$, $10^{8.31}M_{\odot}$ and $10^{8.61}M_{\odot}$ at redshift $0.075$, $0.1$, $0.15$ and $0.2$. Therefore, we choose the stellar mass range of $\rm{pop}_2$ as $[10^{7.9},10^{11.7}]M_{\odot}$ for $z_s<0.2$ and $\bar{n}_2w_{\rm{p}}$ is only calculated at $z_s<0.1$ and $z_s<0.15$ for $M_{*}<10^{8.3}M_{\odot}$ and $M_{*}<10^{8.5}M_{\odot}$ (See Appendix \ref{sec:Z} for a validation). 

The stellar mass limits that DECaLS can reach at $0.2<z_s<0.4$ and $0.5<z_s<0.7$ with $z$ band depth of $22.30$ mag remain to be investigated. We use the DES Year 3 Deep Field catalogs\footnote{https://des.ncsa.illinois.edu/releases/y3a2/Y3deepfields} \citep{2022MNRAS.509.3547H} with photometric redshifts to explore the mass completeness in these redshift ranges. DES Deep can reach a $10\sigma$ $z$ band $2\ arcsec$ source depth of $24.3$ mag, much deeper than DECaLS, and we use the regions with full 8-band $ugrizJHK_s$ coverage ($5.88\ {\rm{deg}}^2$) to get more reliable photoz measurements. We adopt the photoz computed using {\texttt{EAZY}} \citep{2008ApJ...686.1503B} provided by DES and calculate their physical properties using the SED code {\tt{CIGALE}} \citep{2019A&A...622A.103B} with only the $grz$ band fluxes to be consistent with DECaLS. We use the \citet{2003MNRAS.344.1000B} stellar population synthesis models with a \citet{2003PASP..115..763C} initial mass function and a delayed star formation history $\phi(t)\approx t\exp{-t/\tau}$. We adopt three metallicities $Z/Z_{\odot}=0.4,\ 1\ \rm{and}\ 2.5$, where $Z_{\odot}$ is the metallicity of the Sun. And we use the \citet{2000ApJ...533..682C} extinction law for dust reddening with $0<E(B-V)<0.5$. In Figure \ref{fig:fig2} we show the stellar mass - $z$ band magnitude distributions for four redshift ranges $[0.2,0.3]$, $[0.3,0.4]$, $[0.5,0.6]$ and $[0.6,0.7]$. Following \citetalias{2022ApJ...925...31X}, we calculate the $z$ band completeness $C_{95}(M_{*})$ that $95\%$ of the galaxies are brighter than $C_{95}(M_{*})$ in the $z$ band for a given stellar mass $M_{*}$ (red lines). For the DECaLS $z$ band galaxy depth of $22.30$ mag, the complete stellar mass are $10^{9.13}M_{\odot}$, $10^{9.38}M_{\odot}$, $10^{9.87}M_{\odot}$ and $10^{10.06}M_{\odot}$ for the four redshifts. Thus, we choose the stellar mass range of ${\rm{pop}}_2$ as $[10^{9.1},10^{11.9}]M_{\odot}$ and $[10^{9.7},10^{11.9}]M_{\odot}$ for the LOWZ and CMASS redshift ranges respectively, and only measurements at $z_s<0.3$ and $z_s<0.6$ are used for $M_{*}<10^{9.3}M_{\odot}$ and $M_{*}<10^{9.9}M_{\odot}$.

We also split ${\rm{pop}}_2$ into smaller mass bins with an equal logarithmic interval of 0.2 in $\log M_*$ at each redshifts. The final designs are summarised in Table \ref{tab:t1}. 
 
 \subsection{Measurements}
 We adopt the similar method used in \citetalias{2022ApJ...939..104X} to combine the results at different sky regions and redshift bins and to estimate the covariance matrices.
 
 Let $\mathcal{A}=\bar{n}_2w_{\rm{p}}(r_{\rm{p}})$ for better representation. Assuming $\mathcal{A}$ is measured for $N_r$ redshift bins and $N_s$ sky regions (e.g. DECaLS NGC and DECaLS SGC), we further split each region into $N_{\rm{sub}}$ sub-regions for error estimation using jackknife resampling. According to Equation \ref{eq:1}, $\mathcal{A}_{i,j,k}$ can be calculated in $i$th redshift bin, $j$th sky region and $k$th jackknife sub-sample using the Landy–Szalay estimator \citep{1993ApJ...412...64L}. We first combine the measurements at different sky regions using the region area $w_s$ as weight:
 \begin{equation}
     \mathcal{A}_{i,k}=\frac{\sum_{j=1}^{N_{\rm{s}}}\mathcal{A}_{i,j,k}w_{\rm{s},j}}{\sum_{j=1}^{N_{\rm{s}}}w_{\rm{s},j}}\,\,.
\end{equation}
Then, we estimate the mean values and the covariance matrices of the mean values for each redshift bins from $N_{\rm{sub}}$ sub-samples:
\begin{equation}
     \mathcal{A}_i=\sum_{k=1}^{N_{\rm{sub}}}\mathcal{A}_{i,k}/N_{\rm{sub}}\,\,,
\end{equation}
\begin{align}
    C_{ab,i} = \frac{N_{\rm{sub}}-1}{N_{\rm{sub}}}\sum_{k=1}^{N_{\rm{sub}}}(\mathcal{A}_{i,k}(r_{\rm{p}}^a)-\mathcal{A}_{i}(r_{\rm{p}}^a))\\ \notag
    \times(\mathcal{A}_{i,k}(r_{\rm{p}}^b)-\mathcal{A}_{i}(r_{\rm{p}}^b))\,\,,
\end{align}
where $a$ and $b$ denotes the $a$th and $b$th radial bins. Finally, results from different redshift bins are combined
according to the covariance matrices. Let $w_i(r_{\rm{p}})=\sigma^{-1}_i(r_{\rm{p}})$, where $\sigma_i^2(r_{\rm{p}})$ is the diagonal component of $\mathbf{C}_i$,
\begin{equation}
    \mathcal{A}=\frac{\sum_{i=1}^{N_{\rm{r}}}\mathcal{A}_{i}w_i^2}{\sum_{i=1}^{N_{\rm{r}}}w_i^2}\,\,,
\end{equation}
\begin{equation}
    C_{ab} = \frac{\sum_{i=1}^{N_{\rm{r}}}w_i^2(r_{\rm{p}}^a)w_i^2(r_{\rm{p}}^b)C_{ab,i}}{\sum_{i=1}^{N_{\rm{r}}} w_i^2(r_{\rm{p}}^a)\sum_{i=1}^{N_{\rm{r}}} w_i^2(r_{\rm{p}}^b)}
\end{equation}

According to the designs in Table \ref{tab:t1}, we measure $\bar{n}_2w_{\rm{p}}(r_{\rm{p}})$ in the radial range of $0.1h^{-1}{\rm{Mpc}}<r_{{\rm{p}}}<15 h^{-1}\rm{Mpc}$ with $N_{\rm{sub}}=100$ for the 3 redshift ranges. The measurements are extended to scales far beyond the virial radii of halos hosting pop$_1$ galaxies, and thus include information of both centrals and satellites. The results are shown as dots with error bars in Figure \ref{fig:fig3}, \ref{fig:fig4} and \ref{fig:fig5} for the Main sample, LOWZ and CMASS respectively. The square roots of the diagonal components of the covariance matrices are shown as error bars. The measurements are overall good for all mass bins within the whole radial ranges.

\section{Simulation and Modelings} \label{sec:simulation}
In this section, we introduce the simulation and abundance matching method used in this work, and show the constraints on SHMR from our PAC measurements.  
 
\begin{deluxetable*}{cccccccc}
\label{tab:t2}
\tablenum{2}
\tablecaption{Posterior PDFs of the parameters from MCMC for the SHMR models.}
\tablewidth{0pt}
\tablehead{
\colhead{redshift}&\colhead{model}&\colhead{$\log_{10}(M_0)$}&\colhead{$\alpha$}&\colhead{$\delta$}&\colhead{$\beta$}&\colhead{$\log_{10}(k)/\log_{10}(\epsilon)$}&\colhead{$\sigma$}\\
\colhead{}&\colhead{}&\colhead{($\log_{10}(h^{-1}M_{\odot})$)}&\colhead{}&\colhead{}&\colhead{}&\colhead{($\log_{10}(M_{\odot})$)/($\log_{10}(h)$)}&\colhead{}
}
\startdata
$z_s<0.2$&\citetalias{2013ApJ...770...57B}&$11.338^{+0.027}_{-0.028}$&$0.484^{+0.027}_{-0.025}$&$3.041^{+0.134}_{-0.128}$&$1.632^{+0.039}_{-0.041}$&$-1.545^{+0.017}_{-0.017}$&$0.237^{+0.008}_{-0.008}$\\
$0.2<z_s<0.4$&\citetalias{2013ApJ...770...57B}&$11.359^{+0.024}_{-0.025}$&$0.623^{+0.032}_{-0.030}$&$3.248^{+0.131}_{-0.125}$&$1.702^{+0.050}_{-0.054}$&$-1.598^{+0.019}_{-0.020}$&$0.190^{+0.003}_{-0.004}$\\
$0.5<z_s<0.7$&\citetalias{2013ApJ...770...57B}&$11.509^{+0.028}_{-0.029}$&$0.740^{+0.058}_{-0.056}$&$2.964^{+0.159}_{-0.141}$&$2.094^{+0.039}_{-0.039}$&$-1.565^{+0.022}_{-0.025}$&$0.190^{+0.004}_{-0.004}$\\
$z_s<0.2$&DP&$11.732^{+0.020}_{-0.020}$&$0.299^{+0.011}_{-0.012}$&&$1.917^{+0.019}_{-0.020}$&$10.303^{+0.019}_{-0.019}$&$0.233^{+0.008}_{-0.008}$\\
$0.2<z_s<0.4$&DP&$11.579^{+0.012}_{-0.012}$&$0.429^{+0.006}_{-0.006}$&&$2.215^{+0.022}_{-0.022}$&$10.105^{+0.011}_{-0.010}$&$0.201^{+0.003}_{-0.004}$\\
$0.5<z_s<0.7$&DP&$11.624^{+0.010}_{-0.010}$&$0.466^{+0.008}_{-0.008}$&&$2.513^{+0.034}_{-0.033}$&$10.133^{+0.010}_{-0.010}$&$0.192^{+0.004}_{-0.004}$\\
\enddata
\end{deluxetable*}
 
\subsection{CosmicGrowth Simulation}
We use the {\texttt{CosmicGrowth}} simulation \citep{2019SCPMA..6219511J} in this work to model the PAC measurements. The {\texttt{CosmicGrowth}} simulation suite is a grid of high-resolution N-body simulations that are run in different cosmologies using an adaptive parallel P$^3$M code \citep{2002ApJ...574..538J}. We use one of the $\Lambda$CDM simulations with cosmological parameters $\Omega_m = 0.268$, $\Omega_{\Lambda} = 0.732$ and $\sigma_8 = 0.831$ \citep{2013ApJS..208...19H}. The box size is $600\ h^{-1}{\rm{Mpc}}$ with $3072^3$ dark matter particles and softening length $\eta = 0.01\ h^{-1}{\rm{Mpc}}$. Groups are identified using the friends-of-friends (FOF) algorithm with a linking length 0.2 times the mean particle separation. The halos are then processed with HBT+ \citep{2012MNRAS.427.2437H,2018MNRAS.474..604H} to find the subhalos and trace their evolution histories. We use the catalogs of
the snapshots at redshifts of about $0.12$, $0.28$ and $0.57$ to compare with the Main sample, LOWZ and CMASS measurements. Merger timescales of the subhalos with fewer than 20 particles, which may be unresolved, are evaluated using the fitting formula in \citet{2008ApJ...675.1095J} and those that have already merged into central subhalos are abandoned. The halo mass function \citep[see Figure 1]{2019SCPMA..6219511J} and subhalo mass function \citep[see Figure 4]{2022ApJ...925...31X} of the {\texttt{CosmicGrowth}} simulation can be robust down to at least 20 particles ($\sim10^{10.0}h^{-1}M_{\odot}$), which are good enough for this work. 

\begin{figure}
    \plotone{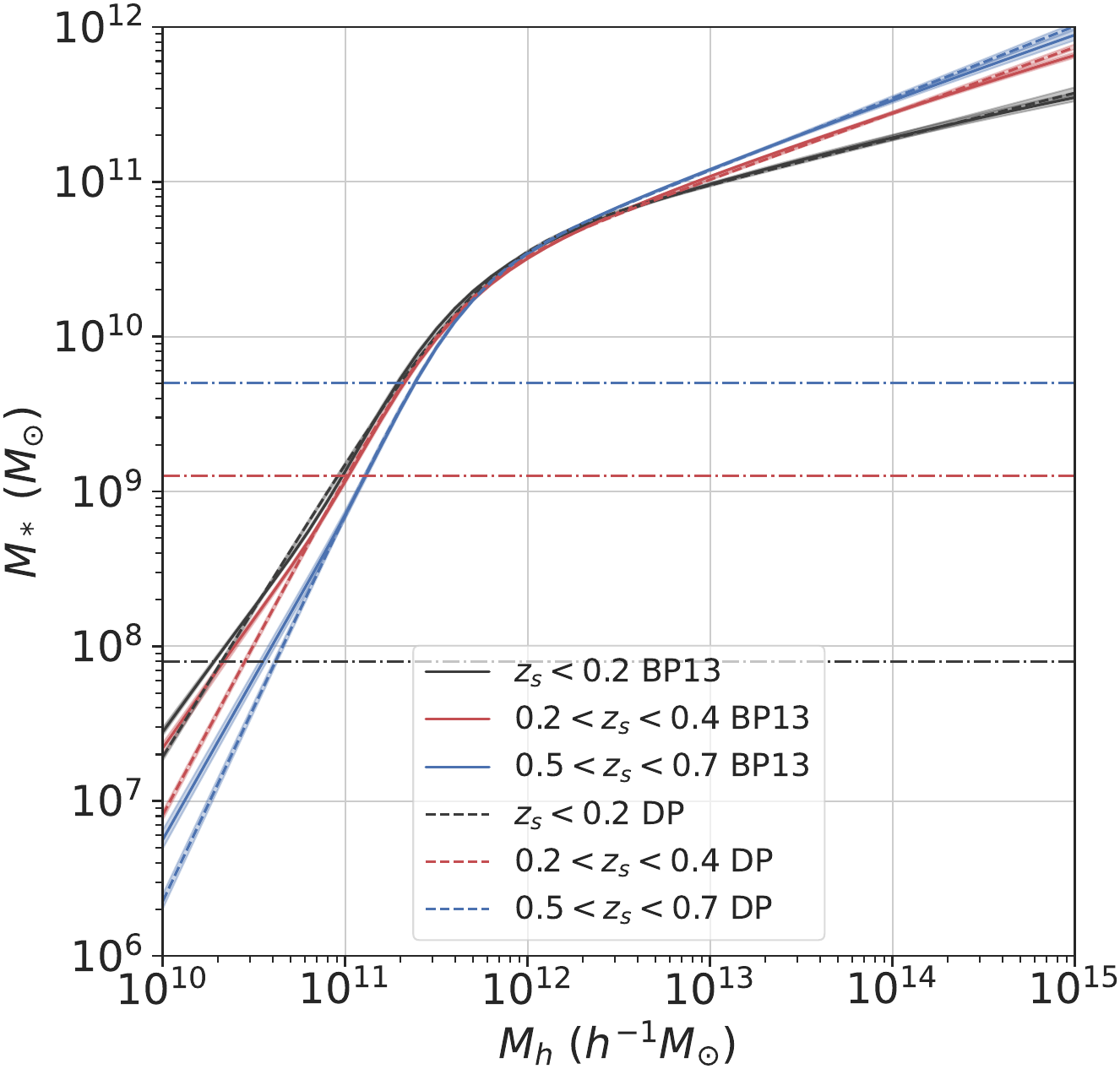}
    \caption{The mean stellar-halo mass relations (lines) and $1\sigma$ errors (shadows) at different redshift ranges and from both the \citetalias{2013ApJ...770...57B} (solid lines) and DP models (dashed lines). The horizontal lines indicate the stellar mass limit covered by the observation data at each redshift matched by color.}
    \label{fig:fig6}
\end{figure}

\begin{figure*}
    \plotone{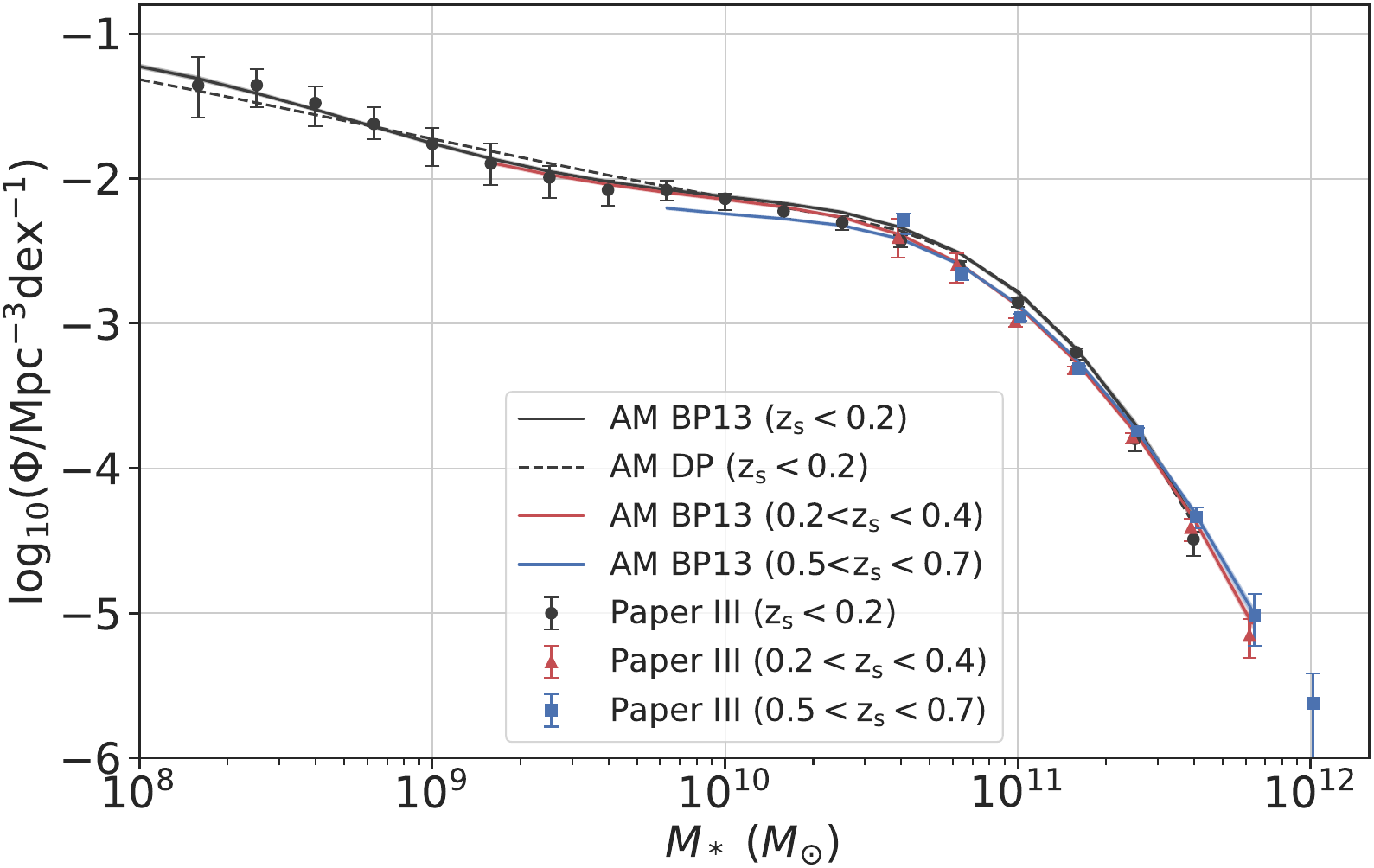}
    \caption{The galaxy stellar mass functions (solid lines) and $1\sigma$ errors (shadows) at different redshift ranges from the \citetalias{2013ApJ...770...57B} model. The results from the DP model at $z_s<0.2$ are also shown in dashed line. The model independent measurements of GSMF from \citetalias{2022ApJ...939..104X} are shown for comparison as dots with error bars.}
    \label{fig:fig7}
\end{figure*}

\begin{figure*}[t]
    \plotone{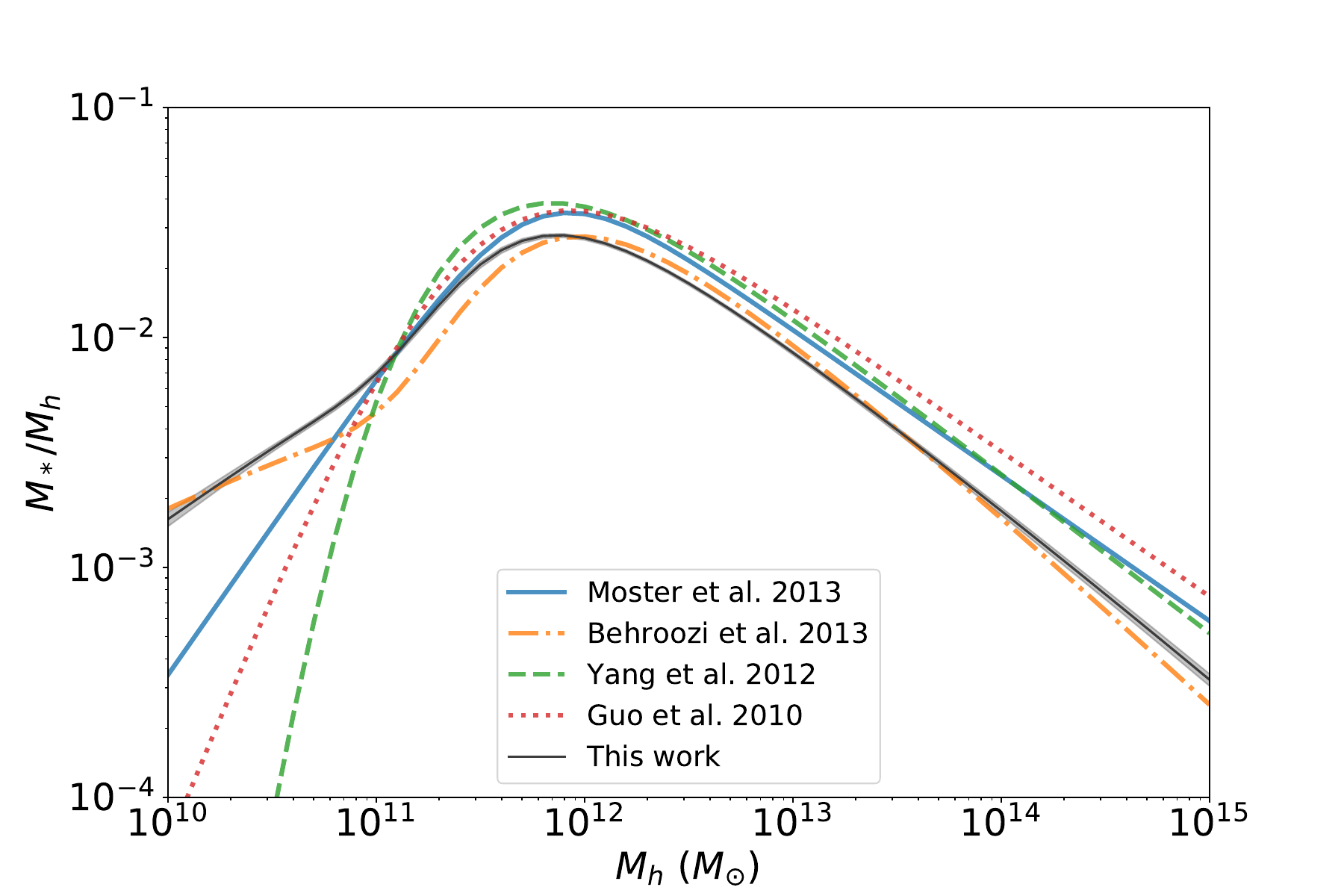}
    \caption{Comparison of the our mean stellar-halo mass function to previous studies at $z_s\sim0.1$. Results compared include those from empirical modeling \citep{2013ApJ...770...57B,2013MNRAS.428.3121M}, from abundance matching \citep{2010MNRAS.404.1111G} and from Conditional Stellar Mass Function (CSMF) modeling \citep{2012ApJ...752...41Y}.}
    \label{fig:fig8}
\end{figure*}

\begin{figure}
    \plotone{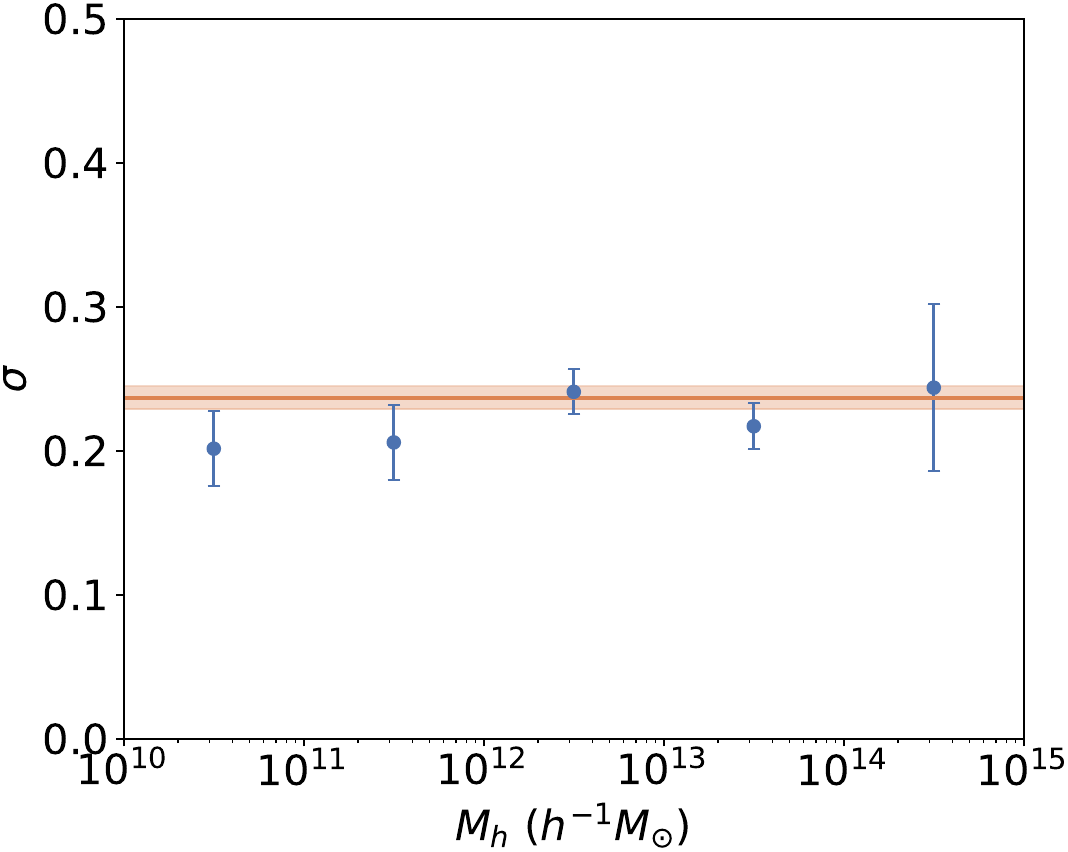}
    \caption{Halo mass dependence of the scatter in the SHMR at $z_s<0.2$. Blue dots with error bars show the results from the multi-$\sigma$ fittings. Orange line with shadow is the previous constraint with a single $\sigma$.}
    \label{fig:fig9}
\end{figure}

\subsection{Subhalo Abundance Matching}
To parameterize the SHMR, the most commonly used five-parameter formula is a double power law with a scatter \citep{2010MNRAS.402.1796W,2012ApJ...752...41Y,2013MNRAS.428.3121M}:
\begin{equation}
    M_{*} = \left[\frac{2k}{(M_{{\rm{acc}}}/{M_0})^{-\alpha}+(M_{{\rm{acc}}}/{M_0})^{-\beta}}\right]\,.\label{eq:8}
\end{equation}
Here we define $M_{{\rm{acc}}}$ as the viral mass $M_{{\rm{vir}}}$ of the halo at the time when the galaxy was last the central dominant object. We use the fitting formula in \citet{1998ApJ...495...80B} to find $M_{{\rm{vir}}}$. The scatter in $\log(M_*)$ at a given $M_{{\rm{acc}}}$ is described with a Gaussian function of the width $\sigma$. We also let $\alpha<\beta$ so that $\alpha$ and $\beta$ represent the slopes of the high and low mass ends of the SHMR respectively.
 
However, \citet{2013ApJ...770...57B} found that the SHMR of the double power law form (hereafter DP) fail to reproduce the upturn feature in the GSMF at $M_{*}<10^{9.5}M_{\odot}$. They provided a six-parameter formula (hereafter \citetalias{2013ApJ...770...57B}) for the SHMR of low mass galaxies:
\begin{align}
\log_{10}(M_*)=\log_{10}(\epsilon M_{0})+f\left(\log_{10}\left(\frac{M_{{\rm{acc}}}}{M_0}\right)\right)-f(0)\\ \notag
f(x)=-\log_{10}(10^{-\beta x}+1)+\delta\frac{(\log_{10}(1+\exp{(x)}))^{\alpha}}{1+\exp{(10^{-x})}} \,, \label{eq:9}
\end{align}
with also a scatter $\sigma$ in $\log(M_*)$. At $M_{{\rm{acc}}}\ll M_{0}$ and $M_{{\rm{acc}}}\gg M_{0}$, this formula degenerate into power laws with indices $\beta$ and $\alpha$.
 
We model the PAC measurements using both the \citetalias{2013ApJ...770...57B} and DP forms for the three redshift ranges. As we will show, both the \citetalias{2013ApJ...770...57B} and DP models are able to fit the measurements in the LOWZ and CMASS equally well, while the \citetalias{2013ApJ...770...57B} model is strongly favored to fit the measurements for $M_*<10^{10}M_\odot$ in the Main sample.

In simulation, the correlation functions are calculated using the tabulated method \citep{2016MNRAS.458.4015Z, 2022ApJ...928...10G} to avoid redundant computation in the fitting process. We define the $\chi^2$ as
\begin{equation}
    \chi^2=\sum_{i=1}^{N_{\rm{m_1}}}\sum_{j=1}^{N_{\rm{m_2}}}(\mathcal{A}^{{\rm{PAC}}}_{i,j}-\mathcal{A}^{{\rm{AM}}}_{i,j})^{T}\mathbf{C}^{-1}_{i,j}(\mathcal{A}^{{\rm{PAC}}}_{i,j}-\mathcal{A}^{{\rm{AM}}}_{i,j})\,\,,
\end{equation}
 where $N_{{\rm_{m_1}}}$ and $N_{{\rm_{m_2}}}$ are the numbers of mass bins of ${\rm{pop}}_1$ and ${\rm{pop}}_2$, $\mathcal{A}^{{\rm{PAC}}}$ and $\mathcal{A}^{{\rm{AM}}}$ are the measurements and model predictions, $\mathbf{C}^{-1}$ is the inverse of the covariance matrix $\mathbf{C}$ and $T$ denotes matrix transposition. We use the Markov chain Monte Carlo (MCMC) sampler {\tt{emcee}} \citep{2013PASP..125..306F} to perform maximum likelihood analyses of $\{M_0,\alpha,\beta,k,\sigma\}$ for the DP model and $\{M_0,\alpha,\delta,\beta,\epsilon,\sigma\}$ for the \citetalias{2013ApJ...770...57B} model.
 
\subsection{Evolution of the Stellar-Halo Mass Relation}
The marginalized posterior PDFs of the parameters are listed in Table \ref{tab:t2} for the 3 redshift ranges and for both the \citetalias{2013ApJ...770...57B} and DP models, and we also show the joint posterior distributions of the parameters in Figure \ref{fig:figa1}-\ref{fig:figa6} using \texttt{corner} \citep{2016JOSS....1...24F}. The corresponding $\bar{n}_2w_{\rm{p}}(r_{\rm{p}})$ from the \citetalias{2013ApJ...770...57B} model in the 3 redshift ranges are shown as lines with shadows in Figure \ref{fig:fig3}, \ref{fig:fig4} and \ref{fig:fig5}. The fitting is good overall for all mass bins in our samples, and all the parameters are constrained well with percent or even sub-percent level errors. We also show the DP model fittings for the Main sample redshift range in Figure \ref{fig:figz1}. We find that the DP model overpredicts the number of galaxies at $M_*=10^{9.5} M_\odot$ while underpredicts at $M_*<10^{8.6} M_\odot$. The \citetalias{2013ApJ...770...57B} model describes the SHMR for small galaxies better than the DP model. 

The best-fit stellar-halo mass relations and the $1\sigma$ errors are shown in Figure \ref{fig:fig6}. Results from the \citetalias{2013ApJ...770...57B} and DP models are shown by solid lines and dashed lines for $z_s<0.2$ (black), $0.2<z_s<0.4$ (red) and $0.5<z_s<0.7$ (blue). We also plot the stellar mass limit covered by the observation data at each redshift with the horizontal lines. In the mass ranges covered by the observation data, all the SHMRs are constrained to percent level, and the results from the \citetalias{2013ApJ...770...57B} and DP models are in good agreement with each other in the LOWZ and CMASS redshift ranges. However, as mentioned above, the DP model is unable to recover the upturn at the low mass end of the SHMR in the Main sample redshift range. 

Our accurate SHMR determination shows that halos with a fixed mass at massive end ($M_h>M_{0}$) host more massive galaxies at higher redshift, which quantifies the downsizing of massive galaxies. On the other hand, our result indicates an opposite trend in the evolution for small halos ($M_h <M_{0}$), which means that small galaxies are continuously forming since $z_s=0.7$,  although the conclusion must be treated with caution as the mass range covered by the observation data is limited for $ M_h<M_0$ at $z_s=0.6$.

We also show the GSMFs in different redshift ranges derived from the \citetalias{2013ApJ...770...57B} model in Figure \ref{fig:fig7}. The GSMFs from our model are all constrained to sub-percent level. The model independent measurements from \citetalias{2022ApJ...939..104X} are also plotted for comparison. We find that the two measurements are in good agreement with each other at all the 3 redshift ranges, proving that both measurements are robust in the whole mass range. At the massive end ($M_*>10^{11.0}M_{\odot}$), the results from \citetalias{2022ApJ...939..104X} were compared with the photoz results from the DESI Legacy Imaging Survey \citep{2021MNRAS.501.3309Z} and we found great consistency (see Figure 5 there). Thus, the 3 independent measurements confirm that our measurements of the GSMF at the massive end are reliable. Our measurements indicate that the GSMF has nearly no evolution since $z_s=0.7$ for $M_*>10^{10.6}M_{\odot}$ and slightly increases with decreasing redshift for smaller stellar mass. We also list the GSMFs from the \citetalias{2013ApJ...770...57B} model in Table \ref{tab:tb1}. We also plot the GSMF at $z_s<0.2$ from the DP model in Figure \ref{fig:fig7}, it is clearly that the DP model fails to capture the upturn in the GSMF and the deviation starts from $10^{10.0}M_{\odot}$. 

\begin{figure*}
    \plotone{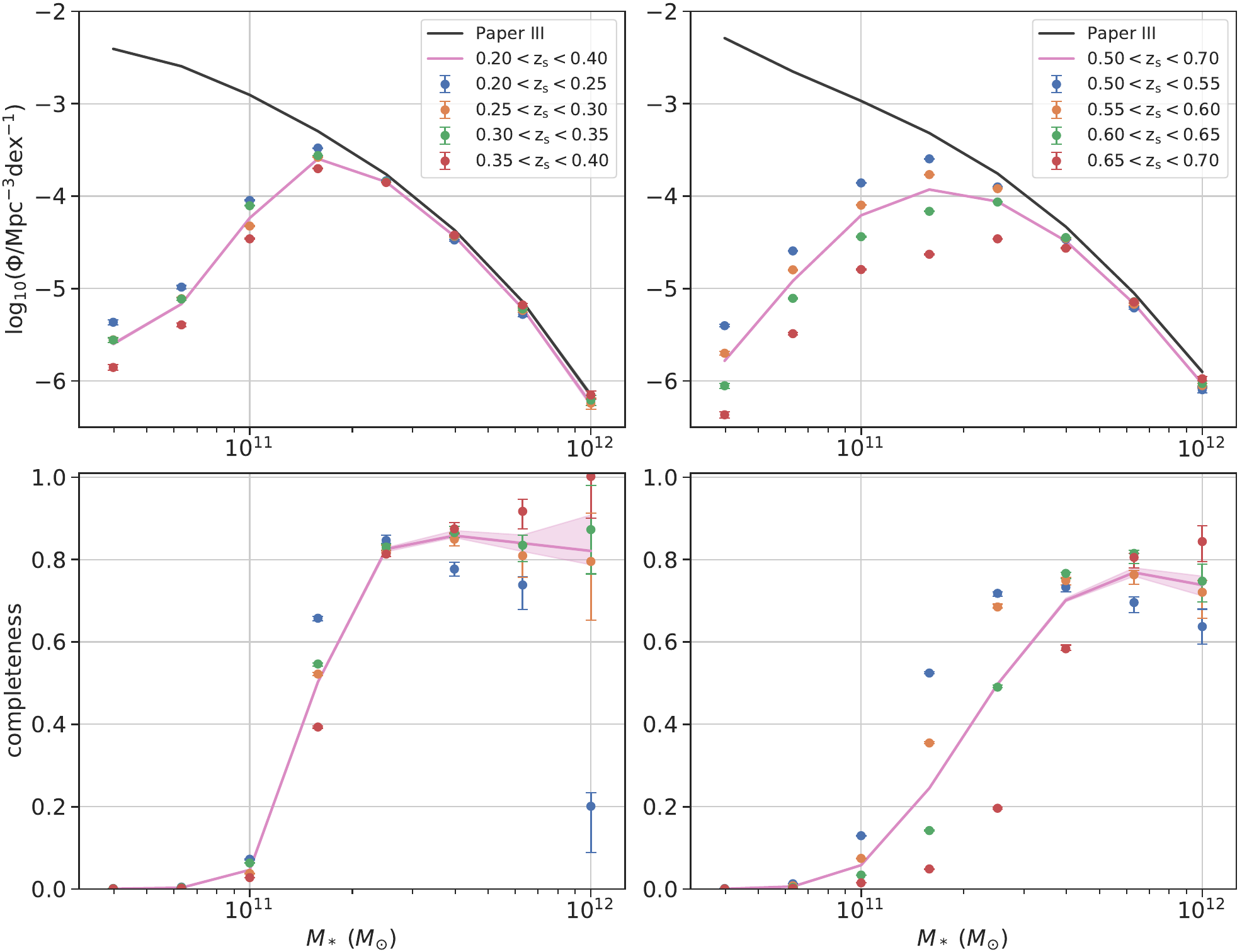}
    \caption{Top: GSMFs of the LOWZ (left) and CMASS (right) samples in different redshift ranges comparing to the DECaLS photoz measurements in \citetalias{2022ApJ...939..104X} ($0.2<z_p<0.4$ and $0.5<z_p<0.7$). Bottom: stellar mass completeness of the LOWZ (left) and CMASS (right) samples.}
    \label{fig:fig10}
\end{figure*}

\begin{figure*}
    \plotone{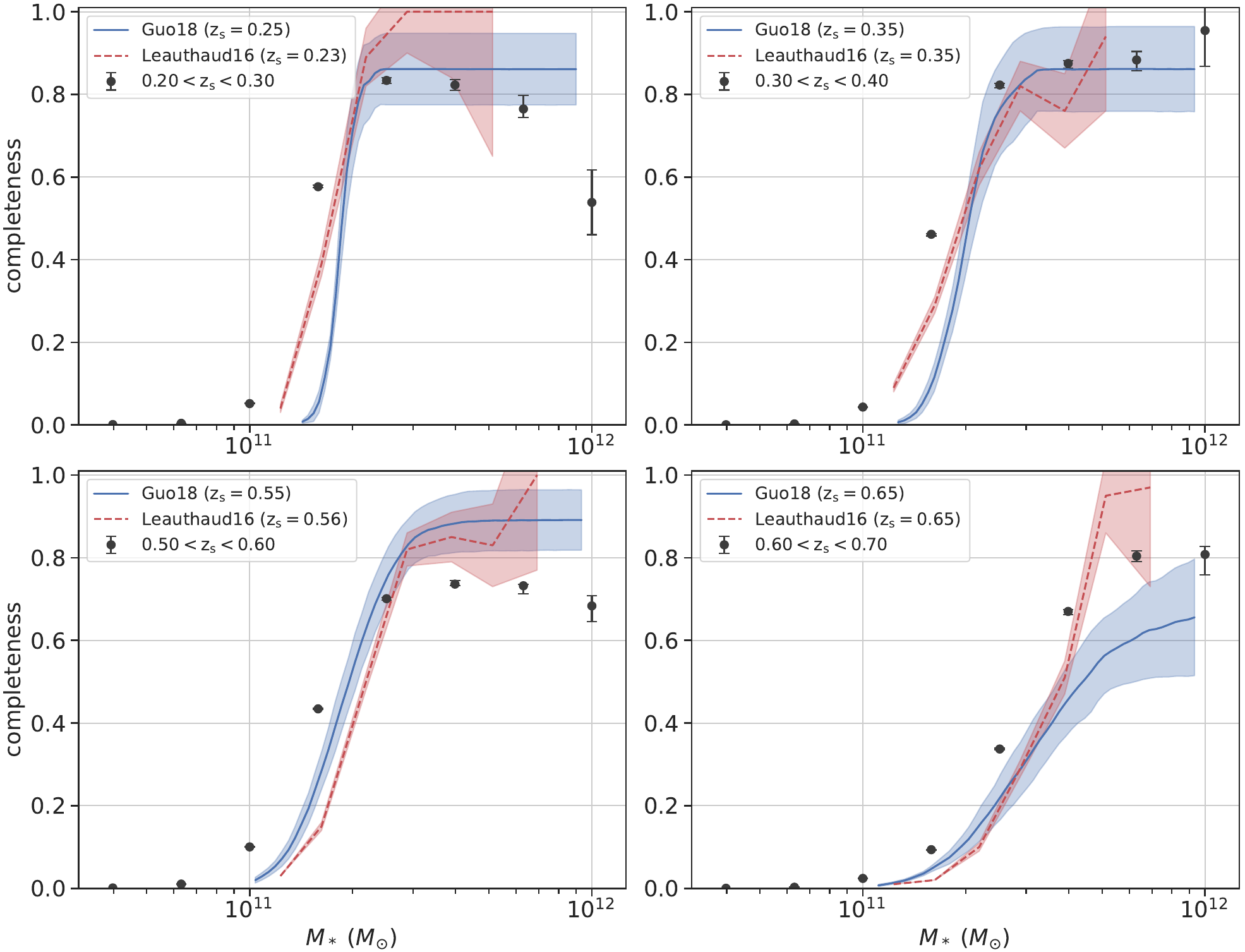}
    \caption{Stellar mass completeness comparing to the results from \citet{2016MNRAS.457.4021L} and \citet{2018ApJ...858...30G}.}
    \label{fig:fig11}
\end{figure*}

Combining the SHMR and GSMF measurements, our results favor the physical picture that massive galaxies are quenched since at least $z_s=0.7$ while their host halos are still assembling their mass, and that low mass galaxies are still forming stars in an efficient way. Our results are inconsistent with a few previous works \citep{2013MNRAS.428.3121M,2013ApJ...770...57B,2019MNRAS.488.3143B} for massive galaxies and halos. In these works, the stellar mass increases with decreasing redshift for fixed halo mass at $M_h>M_{0}$. It may be due to that the GSMFs they used to constrain the SHMR at different redshifts are not well calibrated. Their GSMFs are still increasing with decreasing redshift at the massive ends. Their GSMFs are usually from different surveys that may have systematic bias between each other. Due to the exponentially decreasing feature of the GSMF at the massive end, a very small offset can cause a significant change in the GSMF. As we mentioned in \citetalias{2022ApJ...939..104X} (See Appendix C there), at the high mass end, the GSMF of the north and south parts of the DESI Legacy Imaging Surveys can still have a $30\%$ offset with already a $500\ {\rm{deg}}^2$ footprint overlapped for calibration. Different methods for source extraction, photometry and SED can also introduce further bias. Moreover, the GSMFs they used at high redshifts are usually from the deep spectroscopic or photometric surveys with relatively small survey areas, thus the cosmic variance may become important for the massive end. One of the advantages for measuring the SHMR using PAC is that we can provide measurements in a uniform way, and we can minimize the systematic bias between different redshifts.

In Figure \ref{fig:fig8}, we compare our measurement of SHMR to previous studies at $z_s\sim0.1$. In these works, \citet{2013ApJ...770...57B} and \citet{2013MNRAS.428.3121M} collected the GSMF and/or cosmic star formation rate measurements from different surveys and modeled the evolution of the SHMR using empirical modeling, \citet{2010MNRAS.404.1111G} modeled the SHMR in the local universe with the SDSS DR7 GSMF \citep{2009MNRAS.398.2177L} and galaxy correlations using abundance matching, and \citet{2012ApJ...752...41Y} constrained the evolution of SHMR based on the GSMF and conditional stellar mass function \citep[CSMF,][]{2009ApJ...695..900Y} measurements at different redshifts. The halo masses are all corrected to $M_{{\rm{vir}}}$. However, there are still some differences in the definition of subhalo mass. For satellites, \citet{2013ApJ...770...57B} and \citet{2013MNRAS.428.3121M} use the peak progenitor mass ($M_{{\rm{peak}}}$) while others use the last accretion mass ($M_{\rm{acc}}$). Here we assume that the definition of the subhalo mass has a negligible effect. 

At the high mass end, results from different studies are relatively consistent, since the measurements are robust for this mass range with the large area spectroscopic surveys, though discrepancies still exist that may be due to some systematics in the measurements of the stellar mass. At the low mass end, the discrepancies between studies are large, since the constraints are mainly from the GSMF measurements in SDSS, which have a very limit volume ($z_s<0.03$) for low mass galaxies ($M_{*}<10^{9.0}M_{\odot}$). Instead, making full use of the deep photometric data, we can give a precise measurement of the SHMR at the low mass end with PAC.

\begin{figure*}
    \plotone{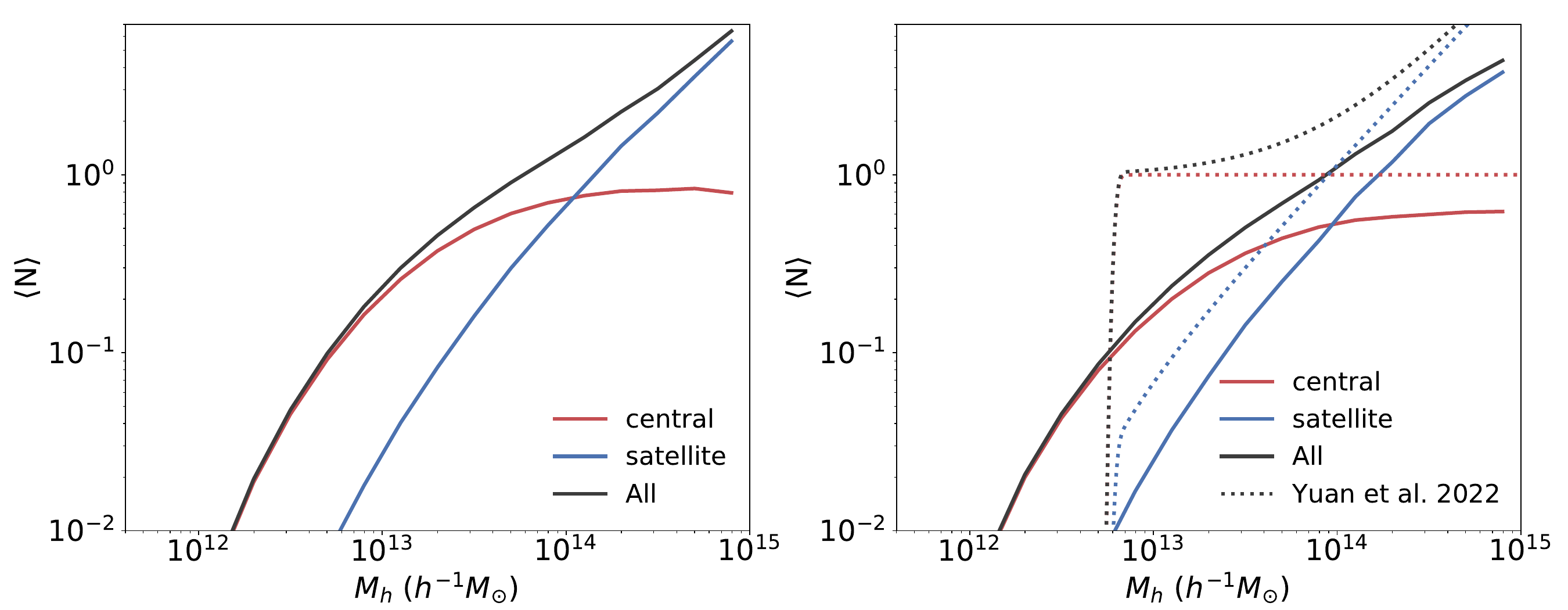}
    \caption{HODs of the LOWZ (left) and CMASS (right) samples derived from our SHMR after taking the stellar mass completeness into account. The HOD of the CMASS sample from \citet{2022MNRAS.515..871Y} are also presented for comparison.}
    \label{fig:fig12}
\end{figure*}

\subsection{Halo Mass Dependence of the Scatter}
The SHMR models we used above adopt a constant scatter $\sigma$. However, whether the scatter depends on halo mass or not is still under debate. Thus, we model the PAC measurements at $z_s<0.2$ again with the \citetalias{2013ApJ...770...57B} model but with 5 different scatters for the halo mass ranges of $[10^{10.0},10^{11.0}]h^{-1}M_{\odot}$, $[10^{11.0},10^{12.0}]h^{-1}M_{\odot}$, $[10^{12.0},10^{13.0}]h^{-1}M_{\odot}$, $[10^{13.0},10^{14.0}]h^{-1}M_{\odot}$ and $[10^{14.0},10^{15.0}]h^{-1}M_{\odot}$, respectively. The results are shown in Figure \ref{fig:fig9}. With the very accurate measurements, we find nearly no dependence of the scatter on halo mass down to $10^{10.0}h^{-1}M_{\odot}$. The scatters are all within $0.2-0.25$ and are consistent with the results from the previous single scatter model. 
 
\section{Halo Occupation Distributions of the LOWZ and CMASS LRG Samples} \label{sec:hod}
The halo occupation distributions (HOD) of the LOWZ and CMASS samples are still under debate \citep{2016MNRAS.457.4021L,2018ApJ...858...30G} due to the stellar mass incompleteness, and the reliability of the HOD can influence cosmology studies such as galaxy-galaxy lensing and redshift space distortion. With the GSMF  measured in \citetalias{2022ApJ...939..104X} and the SHMR measured in this work, we can derive the stellar mass incompleteness and the HODs for the LOWZ and CMASS samples. 

\subsection{Stellar Mass Completeness}
We first calculate the stellar mass completeness for the LOWZ and CMASS samples. We adopt the GSMFs measured in \citetalias{2022ApJ...939..104X} as complete references. For $M_*=10^{12}M_{\odot}$,  as discussed in \citetalias{2022ApJ...939..104X} (see Figure 5), the photoz measurement of GSMFs is more reliable and adopted here, because the number of so massive galaxies is very limited in the survey for the two-point statistics. 

We present the GSMFs of the LOWZ and CMASS samples in different redshift ranges in the top panels of Figure \ref{fig:fig10} along with our GSMFs in \citetalias{2022ApJ...939..104X}. Then we derive the stellar mass completeness in the bottom panels and also list them in Table \ref{tab:tc1}. Errors are all estimated using jackknife-resampling. The completeness in general increases with stellar mass and is peaked at around $10^{11.6}M_{\odot}$. The peak completeness is around $85\%$ for the LOWZ and around $80\%$ for the CMASS. Both the LOWZ and the CMASS samples are not complete even at the highest stellar mass $M_*=10^{11.8}M_{\odot}$ and $10^{12.0}M_{\odot}$, and the incompleteness is even worse at the $10^{12.0}M_{\odot}$ especially at the lower redshift bins ($z_s=0.25$ for the LOWZ and $z_s=0.55$ for the CMASS). In Table \ref{tab:tc2}, we mimic the target selection process for the CMASS sample of $10^{11.8}M_{\odot}$ and $10^{12.0}M_{\odot}$ using the DECaLS photoz sample. We find that target selection for the spectroscopic observations drops out $11.5\%$ and $11.2\%$ of galaxies, and only $89.7\%$ and $81.3\%$ of the selected sources have successful redshift measurements. The two effects combined result in about $79\%$ and $76\%$ completeness, which is consistent with our measurement in Figure \ref{fig:fig10}. Therefor, at least for CMASS, the lower completeness at $10^{12.0}M_{\odot}$ may be due to less spectral identification.

We also compare our stellar mass completeness with previous works \citep{2016MNRAS.457.4021L,2018ApJ...858...30G} in Figure \ref{fig:fig11}. Qualitatively the three studies yield similar incompleteness trends. Quantitatively, the results for LOWZ from the different studies are  consistent with each other, while the differences are larger for CMASS. The reason for the differences is not clear. Their results might be sensitive to the total GSMFs used in \citet{2016MNRAS.457.4021L} and the formula used to model the completeness in \citet{2018ApJ...858...30G}.

\subsection{Halo Occupation Distribution}
Combining with the stellar mass completeness, we can derive the HODs for the LOWZ and CMASS samples from our SHMR. We use the completeness in the redshift ranges of $0.2<z_s<0.4$ and $0.5<z_s<0.7$ for LOWZ and CMASS samples respectively, and assume that the galaxies are randomly selected. The HODs for the LOWZ and CMASS samples are shown in Figure \ref{fig:fig12}. 

As an example, we also compare with a recent HOD for CMASS galaxies (dotted lines in Figure \ref{fig:fig12})
presented by \citet{2022MNRAS.515..871Y} who adopted the standard form of \citet{2007ApJ...667..760Z}. \citet{2022MNRAS.515..871Y} constrained the cosmological parameters and HOD parameters simultaneously using the emulator based on the {\texttt{ABACUSHOD}} framework \citep{2022MNRAS.510.3301Y} in the {\texttt{ABACUSSUMMIT}} simulation suite \citep{2021MNRAS.508.4017M}. We find that both the amplitude and shape of our HOD are different from those of \citet{2022MNRAS.515..871Y}. The difference might be mainly caused by the fact that they did not consider the stellar mass incompleteness. It would be interesting to investigate how much the incompleteness would impact on the determination of cosmological parameters. Cosmological probes, such as galaxy-galaxy lensing and redshift space distortion, are especially sensitive to the HOD or the stellar mass completeness. With the accurately measured SHMR and stellar mass completeness, we will revisit this question in a future work.

\section{Summary} \label{sec:sum}
In this work, using the SDSS Main, LOWZ and CMASS spectroscopic samples and DECaLS photometric data, we measure $\bar{n}_2w_{{\rm{p}}}$ down to stellar mass $10^{8.0}M_{\odot}$, $10^{9.2}M_{\odot}$ and $10^{9.8}M_{\odot}$ for redshift ranges $z_s<0.2$, $0.2<z_s<0.4$ and $0.5<z_s<0.7$, respectively. We model the $\bar{n}_2w_{{\rm{p}}}$ with the abundance matching method in N-body simulation and accurately constrain the evolution of SHMR. We summarize our results as follows:
\begin{itemize}
    \item The parameters of the SHMRs are all well constrained (percent level) for the redshift ranges . According to our results, halos with a fixed halo mass host more massive galaxies at higher redshift for the high mass end ($M_h>M_{0}$), and the trend is reversed at the low mass end ($M_h<M_{0}$). This quantifies the downsizing of massive galaxies since $z_s=0.7$, and indicates that small galaxies are still growing faster than their host halos.
    \item With our precise measurement of $\bar{n}_2w_{{\rm{p}}}$ down to stellar mass $10^{8.0}M_{\odot}$ in the local universe ($z_s<0.2$), we find the form of \citet{2013ApJ...770...57B} describes the SHMR at low mass much better than the double power law form.
    \item Adopting a halo mass dependent scatter of SHMR, we demonstrated that the scatter does not vary with halo mass in the wide mass range of $[10^{10.0},10^{15.0}]h^{-1}M_{\odot}$ at high precision, which supports that a constant scatter assumed in many previous studies is good approximation.
    \item The derived GSMFs from our SHMRs are in perfect agreement with the model independent measurements in \citetalias{2022ApJ...939..104X} at all three redshifts,  but the present study extends the GSMF measurement to lower stellar mass. Our results show that the GSMF has little evolution at the massive end $M_*>10^{10.5} M_{\odot}$ since $z_s=0.7$.  
       \item With the accurate SHMR and GSMF measurements, we calculate the stellar mass completeness and HODs for the LOWZ and CMASS samples. We find that the standard HOD modeling may lead to a biased result without properly taking  into account the stellar mass completeness. Our SHMR and stellar mass completeness measurements will be useful in correctly interpreting the cosmological measurements such as modeling galaxy-galaxy lensing and redshift space distortion based on these samples.
\end{itemize}

With the next generation larger and deeper spectroscopic and photometric surveys such as Dark Energy Spectroscopic Instrument \citep{2016arXiv161100036D}, Legacy Survey of Space and Time \citep{2019ApJ...873..111I} and Euclid \citep{2011arXiv1110.3193L}, we can use the PAC method to explore the galaxy-halo connection to higher redshift and lower mass. 

\begin{acknowledgments}
The work is supported by NSFC (12133006, 11890691, 11621303) and by 111 project No. B20019. We gratefully acknowledge the support of the Key Laboratory for Particle Physics, Astrophysics and Cosmology, Ministry of Education. This work made use of the Gravity Supercomputer at the Department of Astronomy, Shanghai Jiao Tong University.

This publication has made use of data products from the Sloan Digital Sky Survey (SDSS). Funding for SDSS and SDSS-II has been provided by the Alfred P. Sloan Foundation, the Participating Institutions, the National Science Foundation, the U.S. Department of Energy, the National Aeronautics and Space Administration, the Japanese Monbukagakusho, the Max Planck Society, and the Higher Education Funding Council for England.

Funding for SDSS-III has been provided by the Alfred P. Sloan Foundation, the Participating Institutions, the National Science Foundation, and the U.S. Department of Energy Office of Science. The SDSS-III web site is http://www.sdss3.org/. SDSS-III is managed by the Astrophysical Research Consortium for the Participating Institutions of the SDSS-III Collaboration including the University of Arizona, the Brazilian Participation Group, Brookhaven National Laboratory, Carnegie Mellon University, University of Florida, the French Participation Group, the German Participation Group, Harvard University, the Instituto de Astrofisica de Canarias, the Michigan State/Notre Dame/JINA Participation Group, Johns Hopkins University, Lawrence Berkeley National Laboratory, Max Planck Institute for Astrophysics, Max Planck Institute for Extraterrestrial Physics, New Mexico State University, New York University, Ohio State University, Pennsylvania State University, University of Portsmouth, Princeton University, the Spanish Participation Group, University of Tokyo, University of Utah, Vanderbilt University, University of Virginia, University of Washington, and Yale University.

The Legacy Surveys consist of three individual and complementary projects: the Dark Energy Camera Legacy Survey (DECaLS; Proposal ID \#2014B-0404; PIs: David Schlegel and Arjun Dey), the Beijing-Arizona Sky Survey (BASS; NOAO Prop. ID \#2015A-0801; PIs: Zhou Xu and Xiaohui Fan), and the Mayall z-band Legacy Survey (MzLS; Prop. ID \#2016A-0453; PI: Arjun Dey). DECaLS, BASS and MzLS together include data obtained, respectively, at the Blanco telescope, Cerro Tololo Inter-American Observatory, NSF’s NOIRLab; the Bok telescope, Steward Observatory, University of Arizona; and the Mayall telescope, Kitt Peak National Observatory, NOIRLab. The Legacy Surveys project is honored to be permitted to conduct astronomical research on Iolkam Du’ag (Kitt Peak), a mountain with particular significance to the Tohono O’odham Nation.
\end{acknowledgments}

\newpage

\bibliography{sample631}{}
\bibliographystyle{aasjournal}

\appendix
\restartappendixnumbering
\section{verifying the completeness}\label{sec:Z}
\begin{figure}[h]
    \plotone{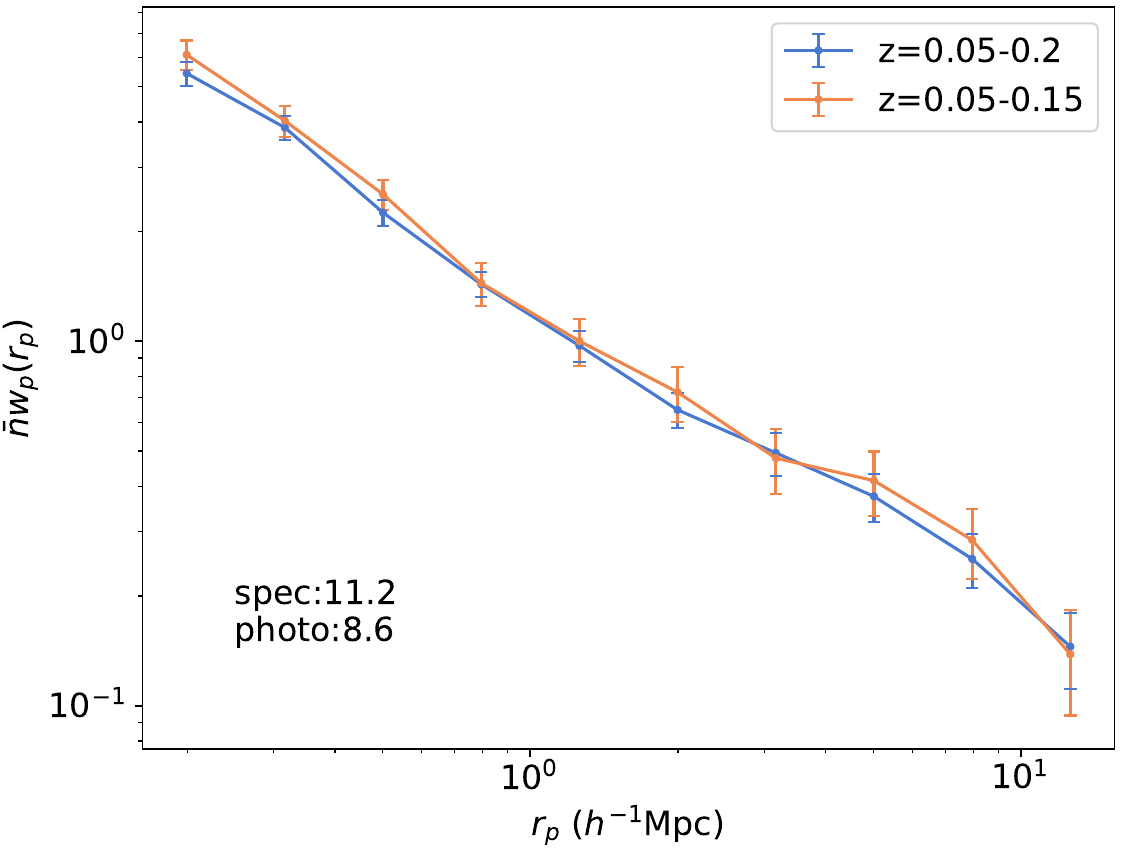}
    \caption{PAC measurements for ${\rm{pop}}_{1}=10^{11.2}M_{\odot}$ and ${\rm{pop}}_{2}=10^{8.6}M_{\odot}$ at $z_s<0.2$ (blue) and $z_s<0.15$ (orange).}
    \label{fig:check}
\end{figure}
We verify our method of determining the mass limits by comparing the $\bar{n}_{2}w_{\rm{p}}$ measurements between $z_s<0.2$ (4 redshift bins) and $z_s<0.15$ (3 redshift bins) for ${\rm{pop}}_{1}=10^{11.2}M_{\odot}$ and ${\rm{pop}}_{2}=10^{8.6}M_{\odot}$. According to our results, for DECaLS, galaxies with $10^{8.6}M_{\odot}$ are exactly complete at $z_s=0.2$ while galaxies with $10^{8.4}M_{\odot}$ are only complete at $z_s=0.15$. If $10^{8.6}M_{\odot}$ is still not complete at $z_s=0.2$, the measured $\bar{n}_{2}w_{\rm{p}}$ from $z_s<0.2$ will be lower than that from $z_s<0.15$. However, as shown in Figure \ref{fig:check}, there is no systematic differences between the results from the two redshift ranges, verifying that $10^{8.6}M_{\odot}$ is complete at $z_s=0.2$ for DECaLS.

\section{Posterior distributions of the parameters}\label{sec:A}
\begin{figure}[h]
    \plotone{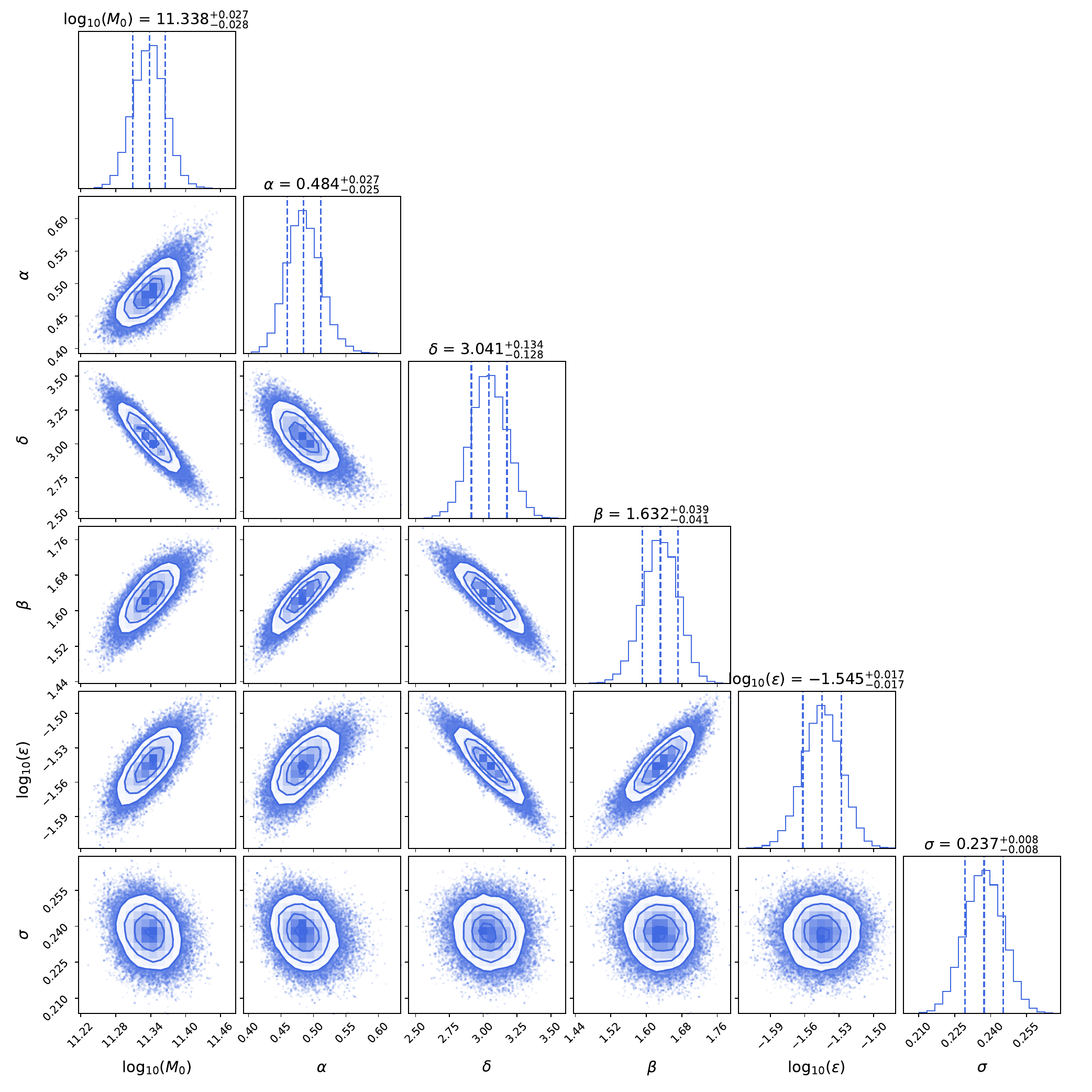}
    \caption{Posterior distributions of the parameters in the \citetalias{2013ApJ...770...57B} model in the Main sample redshift range ($z_s<0.2$). The central value is a median, and the error means $16\sim84$ percentiles after other parameters are marginalized over.}
    \label{fig:figa1}
\end{figure}

\begin{figure}[h]
    \plotone{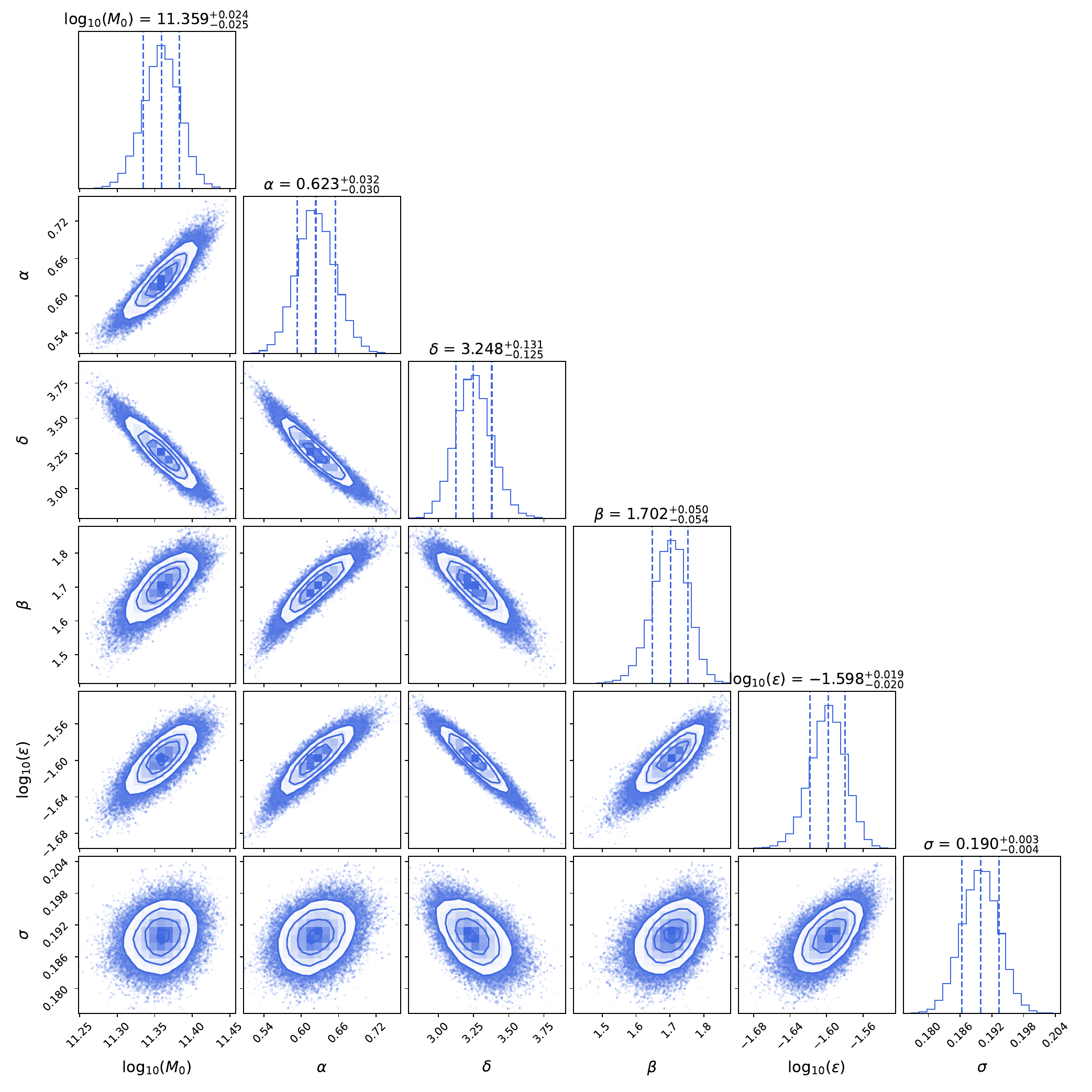}
    \caption{The same as Figure \ref{fig:figa1} but for the \citetalias{2013ApJ...770...57B} model in the LOWZ redshift range ($0.2<z_s<0.4$).}
    \label{fig:figa2}
\end{figure}

\begin{figure}[h]
    \plotone{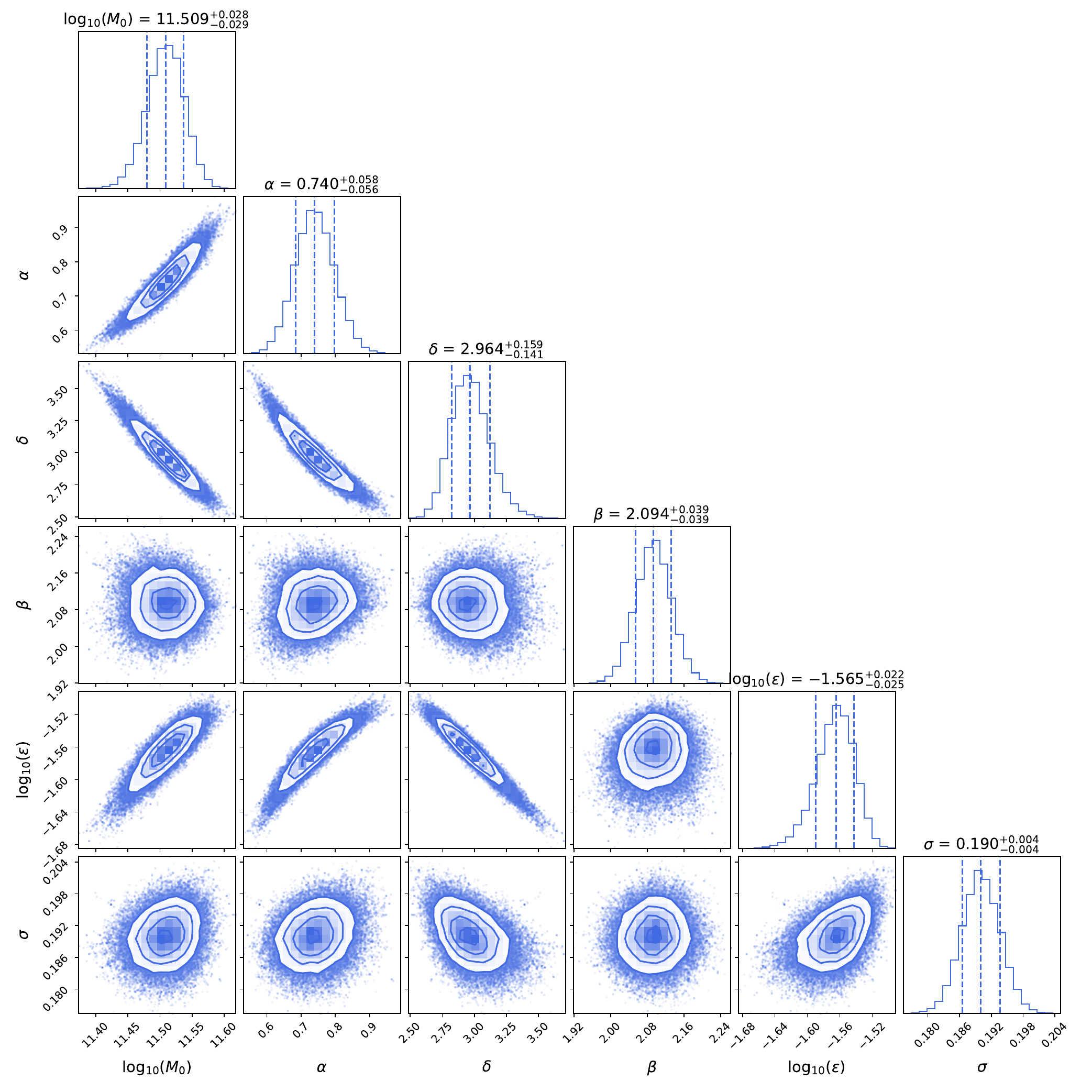}
    \caption{The same as Figure \ref{fig:figa1} but for the \citetalias{2013ApJ...770...57B} model in the CMASS redshift range ($0.5<z_s<0.7$).}
    \label{fig:figa3}
\end{figure}

\begin{figure}[h]
    \plotone{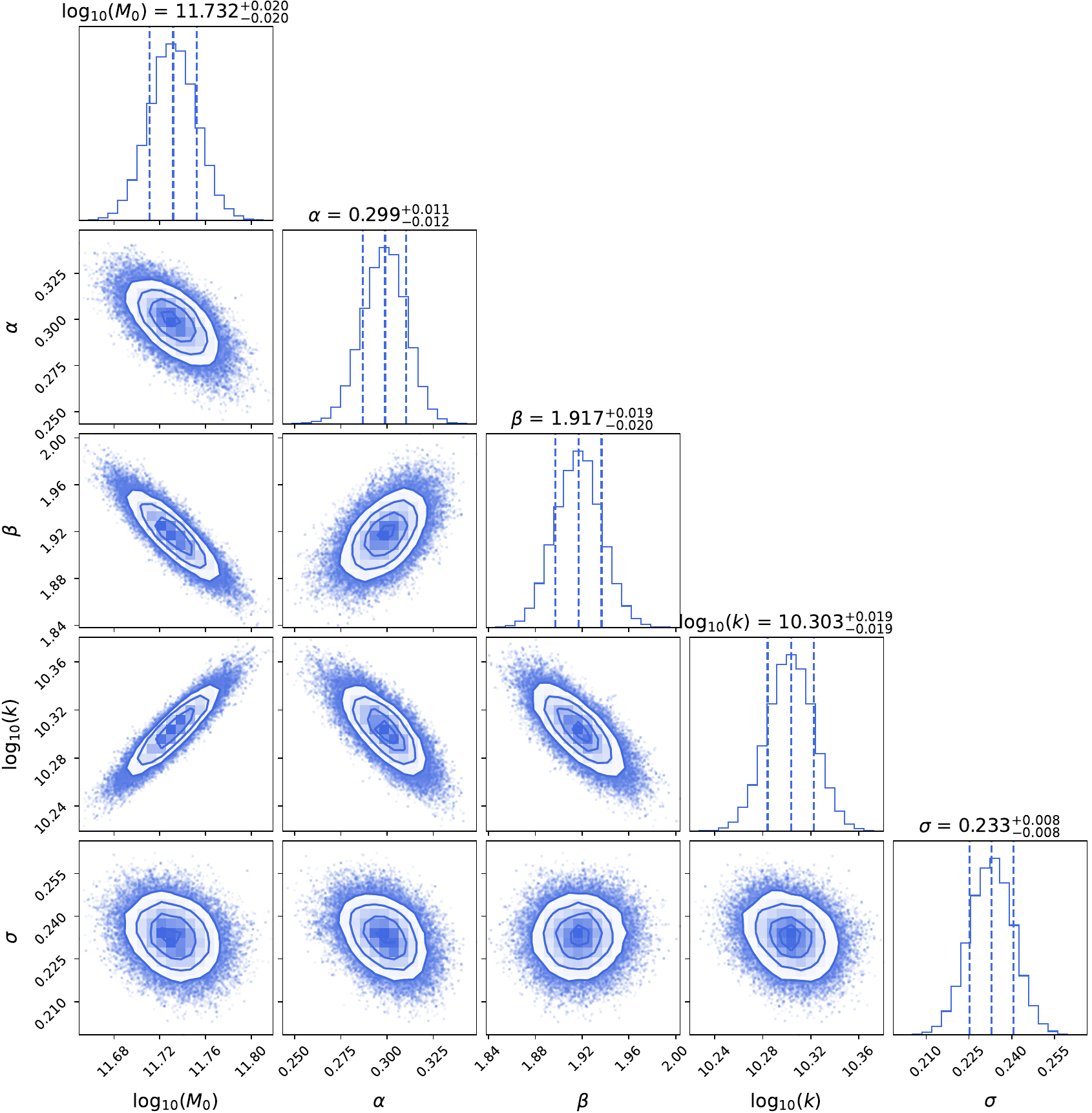}
    \caption{Posterior distributions of the parameters in the DP model in the Main sample redshift range ($z_s<0.2$). The central value is a median, and the error means $16\sim84$ percentiles after other parameters are marginalized over.}
    \label{fig:figa4}
\end{figure}

\begin{figure}[h]
    \plotone{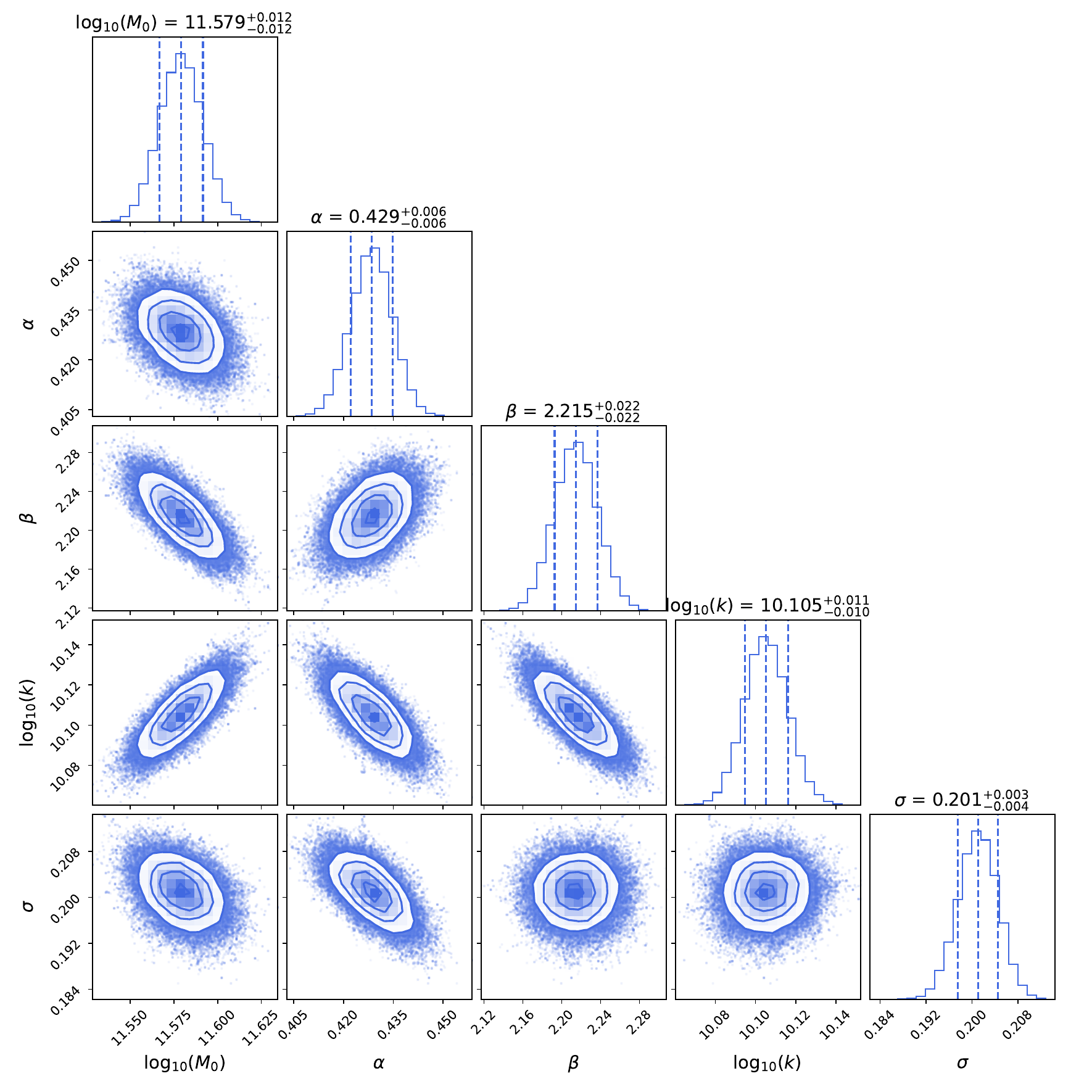}
    \caption{The same as Figure \ref{fig:figa4} but for the DP model in the LOWZ redshift range ($0.2<z_s<0.4$).}
    \label{fig:figa5}
\end{figure}

\begin{figure}[h]
    \plotone{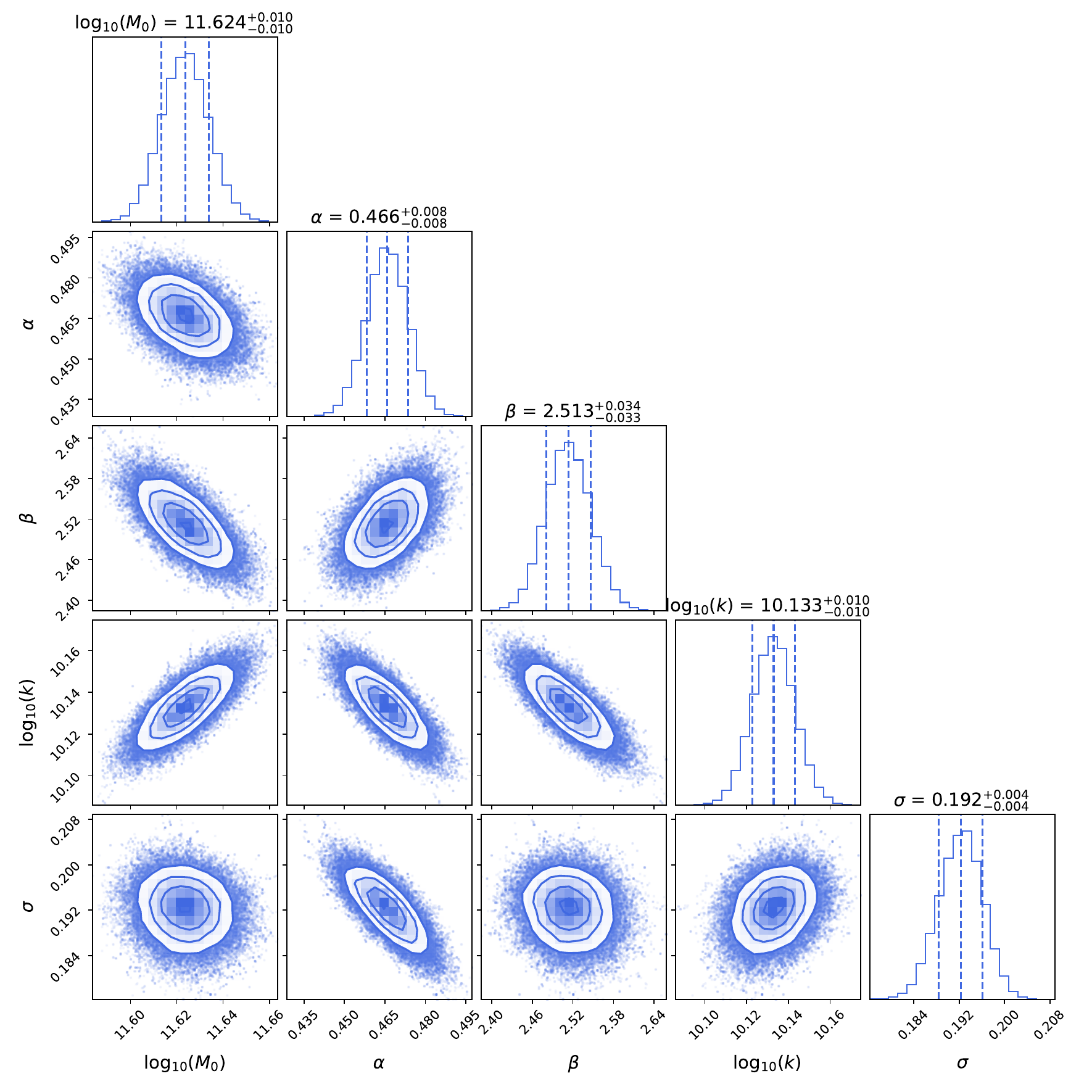}
    \caption{The same as Figure \ref{fig:figa4} but for the DP model in the CMASS redshift range ($0.5<z_s<0.7$).}
    \label{fig:figa6}
\end{figure}

We show the posterior distributions of the parameters of the \citetalias{2013ApJ...770...57B} model at the Main, LOWZ and CMASS redshift ranges in Figure \ref{fig:figa1}, \ref{fig:figa2} and \ref{fig:figa3}. The results of the DP model are also displayed in Figure \ref{fig:figa4}, \ref{fig:figa5} and \ref{fig:figa6}.

\section{The GSMFs from AM in Tabular form }\label{sec:B}
\begin{table}[h]
    \centering
    \caption{The galaxy stellar mass functions $\log_{10}(\Phi/{\rm{Mpc^{-3}dex^{-1})}}$ at different redshifts from the \citetalias{2013ApJ...770...57B} model.}
    \begin{threeparttable}
    \begin{tabular}{cccc}
     \toprule
     $\log_{10}(M_{*}/M_{\odot})$ & $z_s<0.2$ & $0.2<z_s<0.4$ & $0.5<z_s<0.7$ \\
     \midrule
     8.0 & $-1.227_{-0.013}^{+0.012}$ & &\\
     8.2 & $-1.310_{-0.014}^{+0.012}$ & &\\
     8.4 & $-1.412_{-0.011}^{+0.010}$ & &\\
     8.6 & $-1.524_{-0.009}^{+0.008}$ & &\\
     8.8 & $-1.642_{-0.009}^{+0.009}$ & &\\
     9.0 & $-1.758_{-0.010}^{+0.009}$ & &\\
     9.2 & $-1.863_{-0.010}^{+0.008}$ & $-1.890_{-0.008}^{+0.010}$ &\\
     9.4 & $-1.951_{-0.009}^{+0.006}$ & $-1.973_{-0.007}^{+0.009}$ &\\
     9.6 & $-2.022_{-0.009}^{+0.006}$ & $-2.040_{-0.005}^{+0.008}$ &\\
     9.8 & $-2.078_{-0.010}^{+0.006}$ & $-2.096_{-0.005}^{+0.008}$ & $-2.205_{-0.008}^{+0.008}$ \\
     10.0 & $-2.124_{-0.010}^{+0.007}$ & $-2.144_{-0.006}^{+0.009}$ & $-2.243_{-0.007}^{+0.007}$ \\
     10.2 & $-2.170_{-0.010}^{+0.007}$ & $-2.195_{-0.006}^{+0.009}$ & $-2.277_{-0.006}^{+0.007}$ \\
     10.4 & $-2.232_{-0.008}^{+0.007}$ & $-2.268_{-0.007}^{+0.008}$ & $-2.324_{-0.006}^{+0.007}$ \\
     10.6 & $-2.337_{-0.007}^{+0.007}$ & $-2.389_{-0.008}^{+0.008}$ & $-2.417_{-0.007}^{+0.007}$ \\
     10.8 & $-2.513_{-0.008}^{+0.009}$ & $-2.586_{-0.008}^{+0.008}$ & $-2.592_{-0.009}^{+0.007}$ \\
     11.0 & $-2.787_{-0.010}^{+0.011}$ & $-2.878_{-0.008}^{+0.009}$ & $-2.869_{-0.009}^{+0.007}$ \\
     11.2 & $-3.177_{-0.013}^{+0.012}$ & $-3.266_{-0.008}^{+0.011}$ & $-3.247_{-0.010}^{+0.008}$ \\
     11.4 & $-3.693_{-0.020}^{+0.016}$ & $-3.751_{-0.009}^{+0.013}$ & $-3.721_{-0.011}^{+0.011}$ \\
     11.6 & $-4.347_{-0.035}^{+0.026}$ & $-4.343_{-0.014}^{+0.018}$ & $-4.291_{-0.017}^{+0.019}$ \\
     11.8 & & $-5.070_{-0.025}^{+0.028}$ & $-4.979_{-0.031}^{+0.037}$ \\
     \bottomrule
    \end{tabular}  
    \end{threeparttable}
    \label{tab:tb1}
\end{table}

In Table \ref{tab:tb1}, we list the GSMFs inferred from our constrained \citetalias{2013ApJ...770...57B} abundance matching model for the 3 redshift ranges $z_s<0.2$, $0.2<z_s<0.4$ and $0.5<z_s<0.7$. The GSMFs are all constrained to sub-percent level.

\section{stellar mass completeness of the LOWZ and CMASS samples}\label{sec:C}
\begin{table}[h]
    \centering
    \caption{Logarithmic ($\log_{10}$) stellar mass completeness of the LOWZ and CMASS samples at different redshift ranges.}
    \begin{threeparttable}
    \begin{tabular*}{\hsize}{ccccccccc}
     \toprule
     $z_s$ & $10^{10.6}M_{\odot}$ & $10^{10.8}M_{\odot}$ & $10^{11.0}M_{\odot}$ & $10^{11.2}M_{\odot}$ & $10^{11.4}M_{\odot}$ & $10^{11.6}M_{\odot}$\\ & $10^{11.8}M_{\odot}$ & $10^{12.0}M_{\odot}$ \\
     \midrule
     $0.20-0.40$ & $-3.090\pm0.014$ & $-2.496\pm0.008$ & $-1.333\pm0.003$ & $-0.299\pm0.002$ & $-0.083\pm0.003$ & $-0.067\pm0.004$ \\& $-0.076\pm0.010$ & $-0.086\pm0.031$ \\
     $0.20-0.25$ & $-2.862\pm0.029$ & $-2.313\pm0.016$ & $-1.140\pm0.006$ & $-0.182\pm0.003$ & $-0.072\pm0.005$ & $-0.110\pm0.010$ \\& $-0.132\pm0.023$ & $-0.697\pm0.136$ \\
     $0.25-0.30$ & $-3.050\pm0.028$ & $-2.449\pm0.015$ & $-1.422\pm0.008$ & $-0.282\pm0.003$ & $-0.085\pm0.005$ & $-0.071\pm0.008$ \\& $-0.092\pm0.021$ & $-0.099\pm0.070$ \\
     $0.30-0.35$ & $-3.052\pm0.024$ & $-2.446\pm0.014$ & $-1.198\pm0.005$ & $-0.262\pm0.003$ & $-0.081\pm0.004$ & $-0.063\pm0.007$ \\& $-0.079\pm0.016$ & $-0.059\pm0.054$ \\
     $0.35-0.40$ & $-3.338\pm0.029$ & $-2.724\pm0.020$ & $-1.555\pm0.006$ & $-0.405\pm0.003$ & $-0.090\pm0.003$ & $-0.058\pm0.007$ \\& $-0.038\pm0.017$ & $-0.001\pm0.044$ \\
     $0.50-0.70$ & $-3.244\pm0.010$ & $-2.211\pm0.003$ & $-1.237\pm0.002$ & $-0.611\pm0.001$ & $-0.303\pm0.002$ & $-0.154\pm0.003$ \\& $-0.114\pm0.006$ & $-0.132\pm0.014$ \\
     $0.50-0.55$ & $-2.862\pm0.012$ & $-1.888\pm0.007$ & $-0.888\pm0.002$ & $-0.280\pm0.002$ & $-0.144\pm0.003$ & $-0.135\pm0.005$ \\& $-0.158\pm0.012$ & $-0.196\pm0.029$ \\
     $0.55-0.60$ & $-3.159\pm0.017$ & $-2.094\pm0.007$ & $-1.128\pm0.003$ & $-0.450\pm0.002$ & $-0.164\pm0.003$ & $-0.126\pm0.005$ \\& $-0.117\pm0.010$ & $-0.142\pm0.027$ \\
     $0.60-0.65$ & $-3.521\pm0.025$ & $-2.401\pm0.008$ & $-1.468\pm0.004$ & $-0.848\pm0.003$ & $-0.310\pm0.003$ & $-0.116\pm0.004$ \\& $-0.089\pm0.008$ & $-0.126\pm0.026$ \\
     $0.65-0.70$ & $-3.822\pm0.033$ & $-2.786\pm0.013$ & $-1.823\pm0.006$ & $-1.311\pm0.005$ & $-0.707\pm0.004$ & $-0.234\pm0.005$ \\& $-0.094\pm0.009$ & $-0.074\pm0.022$ \\
     \bottomrule
    \end{tabular*}  
    \end{threeparttable}
    \label{tab:tc1}
\end{table}

In Table \ref{tab:tc1}, we list the stellar mass completeness of the LOWZ and CMASS samples at different redshift ranges.

\begin{table}[h]
    \centering
    \caption{Numbers of galaxies with $10^{11.8}M_{\odot}$ and $10^{12.0}M_{\odot}$ after each target selection step of CMASS using DECaLS photoz sample at $0.5<z_p<0.7$.}
    \begin{threeparttable}
    \begin{tabular}{lcc}
     \toprule
     Steps & $10^{11.8}M_{\odot}$ & $10^{12.0}M_{\odot}$ \\
     \midrule
     DECaLS & $10824$ & $1484$\\
     SDSS photometry matched & 10467 & 1403\\
     low-z cut: $|d_{\perp}|>0.55$\tnote{a} & 10320 & 1375\\
     constant mass cut: $i_{{\rm{cmod}}}<19.86+1.6(d_{\perp}-0.8)$ & 10123 & 1364\\
     magnitude limit cut: $17.5<i_{{\rm{cmod}}}<19.9$ & 9953 & 1354\\
     problematic deblending cut: $r_{{\rm{mod}}}-i_{{\rm{mod}}}<2$ & 9868 & 1340\\
     magnitude limit cut: $i_{{\rm{fib2}}}<21.5$\tnote{b} & 9655 & 1328\\
     star–galaxy separation: $i_{{\rm{psf}}}-i_{{\rm{mod}}}>0.2+0.2(20.0-i_{{\rm{mod}}})$ & 9637 & 1327\\
     star–galaxy separation: $z_{{\rm{psf}}}-z_{{\rm{mod}}}>9.125-0.46z_{{\rm{mod}}}$ & 9581 & 1318\\
     CMASS matched\tnote{c} & 8598 (79\%) & 1123 (76\%)\\
     \bottomrule
    \end{tabular}
    \begin{tablenotes}
     \footnotesize
     \item[a] $|d_{\perp}|=(r_{{\rm{mod}}}-i_{{\rm{mod}}})-(g_{{\rm{mod}}}-r_{{\rm{mod}}})/8$
     \item[b] $i_{{\rm{fib2}}}$ is the expected $i$ band magnitude through the SDSS-III 2 arcsec fibres.
     \item[c] Since the footprint of the CMASS sample is slightly smaller than the SDSS photometric sample, the number is corrected using the area ratio.
    \end{tablenotes}
    \end{threeparttable}
    \label{tab:tc2}
\end{table}

In Table \ref{tab:tc2}, we simulate the target selection process of the CMASS sample with $10^{11.8}M_{\odot}$ and $10^{12.0}M_{\odot}$ using the DECaLS photoz sample at $0.5<z_p<0.7$. Notice that the completeness accounts well for the results presented in Figure \ref{fig:fig11}. 



\end{document}